\documentclass[prb,floatfix,twocolumn,amsmath,amssymb,reprint,footinbib,superscriptaddress,showpacs,longbibliography]{revtex4-1}
\usepackage{graphics}      
\usepackage{graphicx}      
\usepackage{longtable}     
\usepackage{url}           
\usepackage{bm}    
\usepackage{braket}        
\usepackage{xcolor}
\usepackage{float}  
\usepackage{natbib}
\usepackage{graphicx}
\usepackage{comment}
\usepackage{siunitx}
\usepackage{appendix}
\usepackage{amsmath,amssymb,epsfig,color}

\usepackage{ulem} 

\usepackage{xcolor}
\usepackage{framed}
\definecolor{shadecolor}{rgb}{0.9,0.9,0.9}

\newcommand{\nn}{\nonumber}
\newcommand{\nl}{\nonumber \\}
\newcommand{\be}{\begin{equation}}
\newcommand{\ee}{\end{equation}}
\newcommand{\bea}{\begin{eqnarray}}
\newcommand{\eea}{\end{eqnarray}}

\newcommand{\tb}[1]{\textbf{\textit{#1}}}

\begin{document}

\title {Phase Space Crystal Vibrations:\\ Chiral Edge States with Preserved Time-reversal Symmetry}

\author{Lingzhen Guo}
\affiliation{Max Planck Institute for the Science of Light, Staudtstrasse 2, 91058 Erlangen, Germany}
\affiliation{School of Science, Nanjing University of Science and Technology, Nanjing 210094, China}

\author{Vittorio Peano}
\affiliation{Max Planck Institute for the Science of Light, Staudtstrasse 2, 91058 Erlangen, Germany}

\author{Florian Marquardt}
\affiliation{Max Planck Institute for the Science of Light, Staudtstrasse 2, 91058 Erlangen, Germany}
\affiliation{Physics Department, University of Erlangen-Nuremberg, Staudtstrasse 5, 91058 Erlangen, Germany}

\begin{abstract}
It was recently discovered that atoms subject to a time-periodic drive can give rise to a crystal structure in phase space. In this work, we point out the atom-atom interactions give rise to collective phonon excitations of phase space crystal via a pairing interaction with intrinsically complex phases that can lead to a phononic Chern insulator, accompanied by topologically robust chiral  transport along the edge of the phase space crystal.  This topological phase is realized even in  scenarios where the time-reversal transformation is a symmetry, which is surprising because  the breaking of time-reversal symmetry  is a strict precondition for topological chiral transport in the standard setting of real space crystals. Our work has also important   implications  for the dynamics of 2D charged particles in a  strong magnetic field. 
\end{abstract}

\date{\today}

\maketitle

\section{Introduction}
The quantum Hall current in a two-dimensional (2D) electron gas  pierced by a magnetic flux is insensitive to weak disorder, because it is carried by chiral edge states that do not have a counter-propagating counterpart \cite{Klitzing1980PRL,Hasan2010RMP}.  With the proposal of the quantum anomalous Hall effect (QAHE) \cite{Haldane1988PRL},  Haldane showed that the key ingredient to engineer chiral edge states is not a net  magnetic field flux,  but rather  the breaking of the time-reversal symmetry (TRS) itself. This insight has opened the way to the current focus on topological transport of neutral excitations such as photons \cite{Haldae_PRL_2008,Lu2014NP,Ozawa_RevModPhys_19}, magnons \cite{Katsura_PRL_2010,shindou2013prb}, phonons \cite{Peano2015PRX,nassar_nonreciprocity_2020} and  cold atoms \cite{Dalibard2011RMP,Goldman2014RPP}.

\begin{figure*}
\centerline{\includegraphics[width=0.95\linewidth]{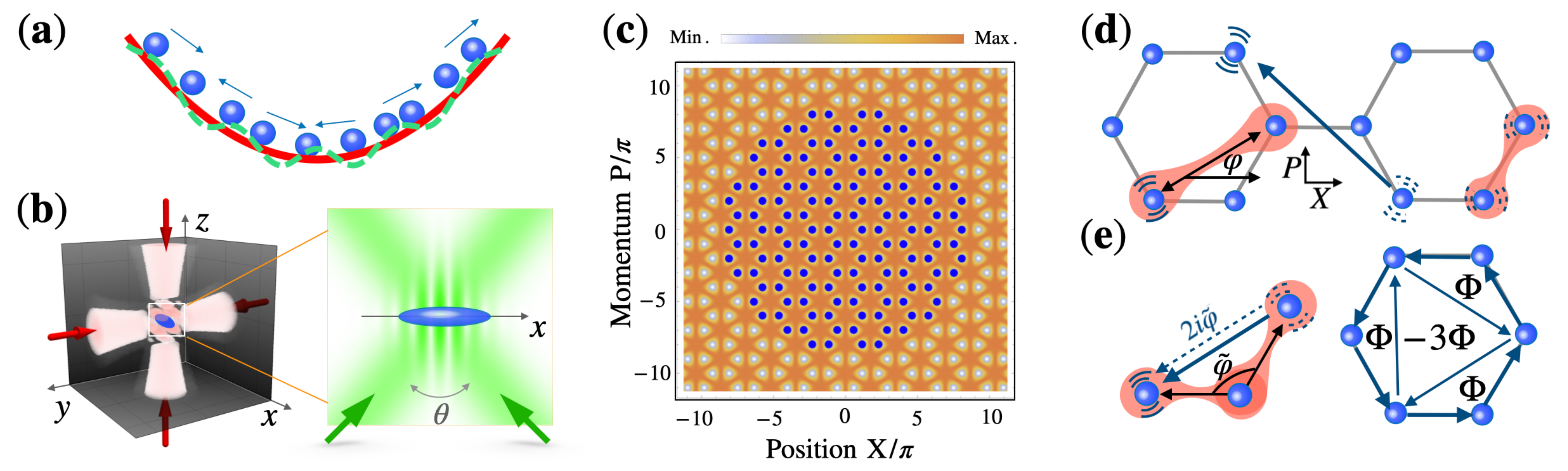}}
\caption{\label{Fig_Honeycomb}
{\bf (a)} Sketch of our model: particles (blue) confined in a static harmonic potential (red) and driven by stroboscopic lattices (green).
{\bf (b)} Experimental implementation of our model with a cloud of cold atoms (blue) confined in quasi-1D tubes by two orthogonally polarized lasers (pink) and subject to a tuneable stroboscopic lattice formed by two lasers (green) intersecting at angle $\theta$. {\bf (c)} Density plot of single-particle Hamiltonian $H_s(X,P)$, with $X, P$ in a rotating frame. The honeycomb lattice sites are occupied by atoms (blue), forming a disk-shaped crystal in phase space. {\bf (d)} Sketch of coupling between the atoms' vibrations. The arrow indicates hopping of a vibrational excitation.  The pink dumbbell-like shapes indicate pairwise creation (annihilation) of excitations.  {\bf (e)}
Sketch of the  pairing-interaction-induced hopping (left panel) and the resulting effective magnetic field (right panel). The induced non-reciprocal hopping pathway  (dashed arrow)   interferes with a direct hopping pathway (solid arrow), leading to a weak staggered magnetic flux of order $\sim\gamma/\omega_0$, see more discussion in Appendix \ref{App:PISMF}.}
\end{figure*}

In this work, we introduce a conceptually different way to generate robust chiral phonon transport, without requiring a breaking of the TRS. A \textit{single quantum} particle moving on a closed path in phase space acquires a Berry phase \cite{Guo2013PRL,Zhang2017pra,Liang2018NJP,Lorch_2019}. Therefore, when one considers the quantum motion in an extended periodic phase-space potential, the resulting matter wave bandstructure supports non-trivial Chern numbers \cite{Leboeuf1990PRL,Leboeuf1992Chaos,Liang2018NJP}. 
However, so far it was unclear how this could lead to the most important consequence of such Chern numbers, namely chiral edge channels, since it is not straightforward to produce a boundary of such a phase-space potential. 
Here we consider a different scenario, demonstrating a new mechanism that occurs already in the \textit{classical} domain, but only if one considers an \textit{interacting many-particle} system. When the particles arrange themselves in the extrema of the underlying phase-space potential, it gives rise to a  so-called phase space crystal \cite{Liang2018NJP}. We will show that the phase-space crystal vibrations acquire a topologically non-trivial band structure, due to  the combination of interactions and symplectic phase-space geometry. In contrast to the single-particle case, the boundary  of the finitely extended phase space crystal naturally supports chiral edge channels in phase space, which are now of phononic (collective) nature. We show that these chiral edge channels can arise even when the driving preserves the time-reversal symmetry. Finally, we explore the implications of our work for the crystal phase of 2D charged particles confined in the Lowest Landau Level (LLL) in the presence of a strong magnetic field.

Similar to Floquet approaches for topology \cite{Lindner2011NP,Rudner2013PRX}, we will be employing time-dependent driving, but in our case the drive itself need not break TRS to generate chiral transport, as a consequence of the nonlocal nature of the time reversal operation in phase space. Our construction produces topological channels in a 2D phase space, starting from 1D real space. This is reminiscent of synthetic dimensions \cite{Celi2014PRL,Ozawa2019NRP,Yuan2018OSA}, but unlike that concept, no extra controlled degree of freedom is needed, we automatically get chiral motion, and besides in synthetic dimensions it is very challenging to create non-trivial lattice structures \cite{Suszalski2016pra} or local and isotropic interactions \cite{Ozawa2019NRP}, in contrast to the present approach.

The remaining part of this paper is organised as follows. In Sec.~\ref{model},  we introduce a model for a kicked harmonic oscillator. Our model  allows to engineer any arbitrary lattice Hamiltonian in phase space by selecting an appropriate kicking sequence. We  present the time-reversal-invariant kicking sequence that generates a honeycomb lattice Hamiltonian in phase space. In Sec.~\ref{PSI}, we investigate the atom-atom interactions for an ensemble of atoms confined in our phase-space lattice.
Starting from the full phase-space many-body Hamiltonian, we derive an effective quadratic Hamiltonian  for the small vibrations about a phase-space crystal equilibrium configuration. 
In Sec.~\ref{TP}, we study the  resulting bulk band structure  and show that it supports a topological band gap with a non-zero Chern number, even though our model is symmetric under time reversal transformation.
In Sec.~\ref{sec:TT}, we investigate the phase space crystal vibrations in the presence of  physical boundaries. We compare our phase-space Chern insulator with standard real space Chern insulators and explain why the breaking of the time-reversal symmetry is not a precondition for robust chiral transport in phase space. In Sec.~\ref{sec:TopTran}, we show that the  disk-shaped phase space crystal of finite size supports chiral edge states and discuss how to observe the ensuing robust transport  in a cold-atom experiment.    In Sec.~\ref{real-space-motion}, we explore the connection between our phase-space dynamics and  an important condensed matter scenario: the 2D real space dynamics of charged particles in an intense magnetic field. In particular, we  discuss  the implications of  our work for the dynamics of the guiding centers of an ensemble of particles frozen in the Lowest-Landau-Level (LLL). In Sec.~\ref{outlook}, we  summarise our results and give an overview of  possible future research directions.

\section{Phase Space Lattice}\label{model}

\subsection{Honeycomb lattice in phase space}

Consider cold atoms trapped in a quasi-1D elongated harmonic potential with the axial trapping frequency $\omega_{ax}$ and transverse trapping frequency $\omega_{tr}\gg \omega_{ax}$. The atoms are driven stroboscopically by multiple optical lattices with kicking frequency $\omega_{ax}/\tau$, where $\tau$ is the dimensionless kicking period. We sketch our model and the experimental implementation in Fig.~\ref{Fig_Honeycomb}(a)-(b). 
The single-atom Hamiltonian is given by
\begin{equation}\label{eq:dimensionlessKHO}
\tilde{H}_s=\frac12(x^2+p^2)+\sum_{n\in \mathbb{Z}}\sum_{q} K_q\cos (k_qx-\phi_q)\delta\Big(\frac{t}{\tau}-\theta_q-n\Big).
\end{equation}
Here, we have rescaled the coordinate, momentum, time and energy by $k^{-1}$, $m\omega_{ax}/k$, $\omega_{ax}^{-1}$ and $m\omega^2_{ax}/k^2$, respectively, where $m$ is the mass and $k$ is the typical  wave vector of the stroboscopic lattices. This model is a generalization of the well-known
kicked harmonic oscillator \cite{Zaslavsky1986ZETF,Moore1995PRL,Chabe2008PRL,Manai2015PRL,Lemos2012NC,Zaslavsky2008Book} in that it allows for pulses with different wavelengths $k_q$.  

For weak near-resonant driving  ($|K_q|\ll 1$, $\tau/2\pi\approx 1$), the single-particle dynamics is dominated by fast harmonic oscillations with slowly changing quadratures $(X,P)$: 
\bea\label{eq-main-xpXP}
\left\{\begin{array}{l}
x(t)=P\sin \big(\frac{2\pi t}{\tau}\big)+X\cos \big(\frac{2\pi t}{\tau}\big)
 \\
p(t)=P\cos \big(\frac{2\pi t}{\tau}\big)-X\sin \big(\frac{2\pi t}{\tau}\big). 
\end{array}\right.\ \ 
\eea
Within rotating-wave approximation (RWA) this leads to a time-independent Hamiltonian (see Appendix \ref{App:PSLH} for detailed derivation)
\bea\label{eq:HsXiPi}
&&H_s(X,P)
=\frac{1}{2}\delta\omega\left(P^2+X^2\right)+\nonumber\\
&&\sum_{q}K_q\cos \Big[k_q\Big(P\sin 2\pi\theta_q+X\cos 2\pi\theta_q\Big)-\phi_q\Big],
\eea
where  $\delta\omega\equiv1-2\pi/\tau$ is the detuning between driving and harmonic frequencies.
 The  extrema of  $H_s(X,P)$ represent stable points.  In the standard kicked harmonic oscillator \cite{Zaslavsky1986ZETF,Moore1995PRL,Chabe2008PRL,Manai2015PRL,Lemos2012NC,Zaslavsky2008Book} and for an appropriate choice of the parameters, the stable points can form a square or triangular bravais lattice. Our more general model allows for more flexibility in the design of 
$H_s(X,P)$ and, in particular, to realize any arbitrary phase-space lattice (see more details in Appendix \ref{App:PSLH}). 

\begin{figure}
\center
\includegraphics[scale=0.25]{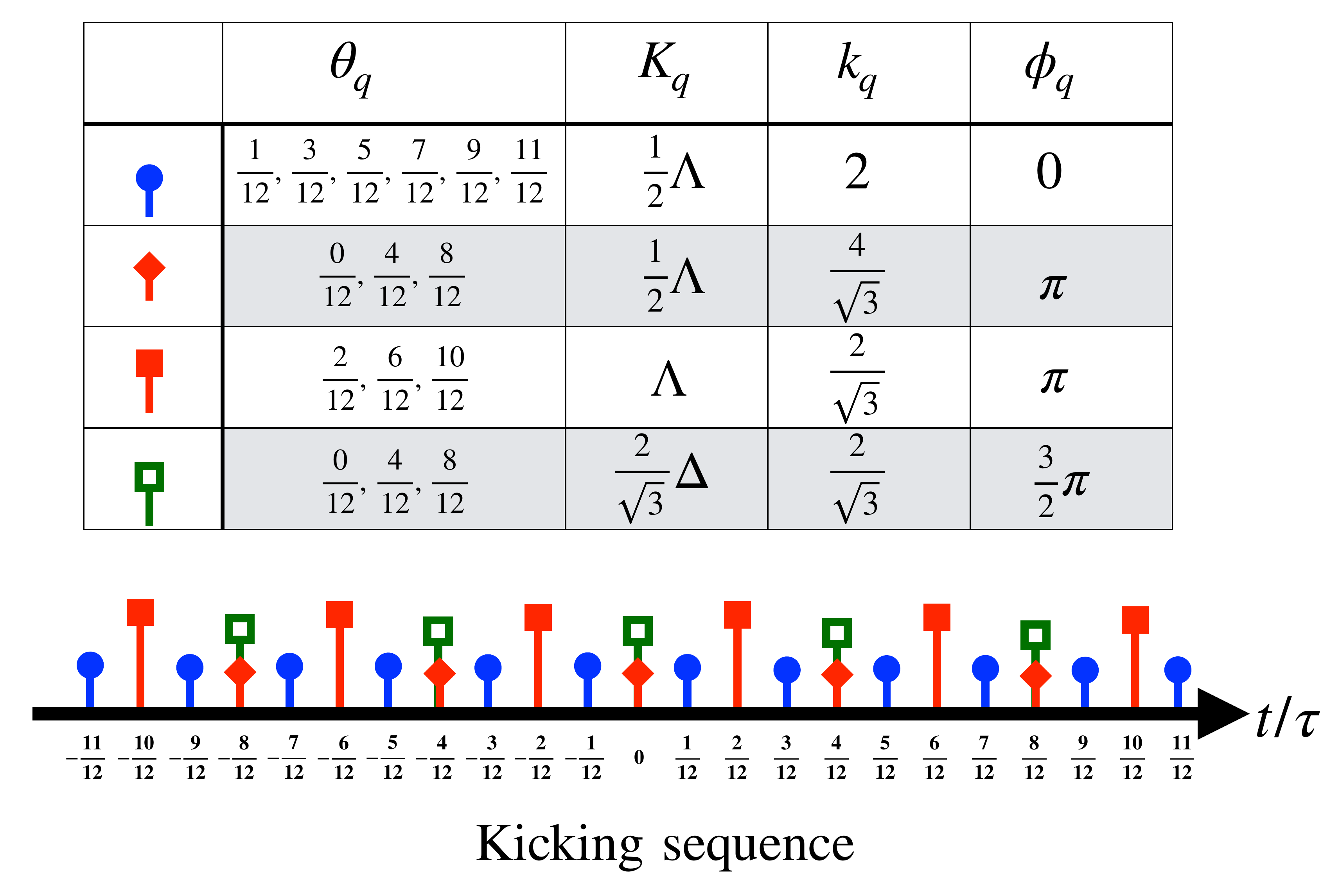}
\caption{Kicking protocol for the honeycomb lattice: (Top) Kicking parameters table and (Bottom) kicking sequence. The kicks listed in the first three lines  lead to the parity-symmetric  Honeycom lattice Hamiltonian, i.e., second term in Eq.~(\ref{PSL}). The kicks in the fourth line induce the parity-symmetry breaking Hamiltonian, i.e., third term in  Eq.~(\ref{PSL}).
}\label{Fig-SM-kicking}
\end{figure}

For concreteness, in the remainder of the paper,  we focus on a scenario where   $H_s(X,P)$  supports a honeycomb lattice of stable points.
By superposing twelve equally spaced kicks as shown in Fig.~\ref{Fig-SM-kicking}, we have the following single-particle Hamiltonian (up to a constant $3\Lambda/2$),
\bea\label{PSL}
H_s(\tb Z)&=&\frac{1}{2}\delta\omega|\tb Z|^2-\frac{16}{9}\omega_0\prod_{n=1}^3\sin^2\big(\frac{1}{2}\tb v_n\cdot\tb Z\big)\nl
&&-\frac{2\Delta}{\sqrt{3}}\sum_{n=1}^3\sin\big(\tb v_n\cdot\tb Z\big).
\eea
Here, we have defined the   vector ${\tb Z}\equiv (X,P)$, and three ancillary vectors $\tb v_1=(\frac{2\sqrt{3}}{3},0)$, $\tb v_2=(-\frac{\sqrt{3}}{3},1)$, $\tb v_3=(-\frac{\sqrt{3}}{3},-1)$. For  $\delta\omega=0$, the minima of $H_s(\tb Z)$ form a honeycomb lattice, cf. Fig.~\ref{Fig_Honeycomb}(c).  
The frequency of small vibrations about these  stable points $\omega_0=-9\Lambda$ can be obtained by the linear expansion of Eq.~(\ref{PSL}). 
The last term in the Hamiltonian~(\ref{PSL})  breaks the inversion symmetry of the honeycomb lattice and, thus, allows for different onsite energies of the two sublattices $\omega_L=\omega_0\pm 2\Delta$.
We note  that $[X,P]=i\lambda$ with the effective Planck constant $\lambda=\hbar k^2/m\omega_{ax}$, which is twice the square of the Lamb-Dicke parameter\cite{Wineland1998}. The condition
$\lambda\ll 1$ ensures that the quantum fluctuations are small
compared to the phase-space distance between neighboring stable points. In the reminder of this paper, we will focus on this parameter regime.

\subsection{Symmetries in phase space}\label{App:TRS}
Next, we analyze  how the time and space symmetries of the full lab-frame Hamiltonian  Eq.~(\ref{eq:dimensionlessKHO}) are reflected into the phase-space symmetries of the RWA Hamiltonian Eq.~(\ref{eq:HsXiPi}).
For concreteness, we focus on the special case of our honeycomb lattice Hamiltonian Eq.~(\ref{PSL}). For the train of pulses displayed in the parameter table in Fig.~\ref{Fig-SM-kicking}, the full Hamiltonian Eq.~(\ref{eq:dimensionlessKHO}) has period $\tau/3$, $\tilde{H}_s(t+\tau/3)=\tilde{H}_s(t)$. This is one-third of the rotation period  $\tau$ of the frame of reference  Eq.~(\ref{eq-main-xpXP}) in which  the RWA Hamiltonian is defined. This discrete time-translational symmetry leads to a three-fold rotational symmetry of the Hamiltonian Eq.~(\ref{PSL}) in phase space.

In addition, the Hamiltonian Eq.~(\ref{eq:dimensionlessKHO}) has time-reversal symmetry $\tilde{H}_s(-t)=\tilde{H}_s(t)$, cf. the kicking sequence displayed below the parameter table in Fig.~\ref{Fig-SM-kicking}. The time-reversal symmetry leads to a mirror-symmetry in phase space $H_s(X,-P)=H_s(X,P)$. Because of the tree-fold rotational symmetry, the Hamiltonian $H_s(X,P)$ has actually three different mirror planes. Each such plane corresponds to two difference reference time for the time-reversal symmetry $\tilde{H}_s(-t+n\tau/3)=\tilde{H}_s(t+n\tau/3),$ and $\tilde{H}_s(-t+n\tau/3+\tau/2)=\tilde{H}_s(t+n\tau/3+\tau/2),$ with $n=0,1,2$ for the three different planes.

For the special case where the pulses in the last row of the table in Fig.~2 have zero amplitude (corresponding to $\Delta=0$), the Hamiltonian Eq.~(\ref{eq:dimensionlessKHO}) has also parity symmetry $\tilde{H}_s(x)=\tilde{H}_s(-x)$. The parity symmetry leads to an additional mirror plane in phase space, $H_s(-X,P)=H_s(X,P)$. Combined with the time-reversal symmetry $H_s(X,-P)=H_s(X,P)$, it leads to the two-fold rotational symmetry, $H_s(-X,-P)=H_s(X,P)$. Thus,  our phase space crystal (for $\Delta=0$) has the full point-group symmetry ${\cal C}_{6\nu}$ (six-fold rotations and six mirror planes) of the underlying triangular Bravais lattice.

\section{Phase Space Crystal}\label{PSI}

\subsection{Phase space interaction}\label{subsecMBD}
The interaction of neutral cold atoms in a tight 1D-trap is captured by an effective two-body contact potential\cite{Bloch2008RMP}, $V(x_i-x_j)=\gamma\delta(x_i-x_j)$. Since atoms that are localized about distant phase-space points will still collide in the course of their lab-frame trajectories, the lab-frame contact interaction gives rise to an effective  long-range interaction in the rotating frame \cite{sacha2015scirep,sacha2015pra,guo2016pra,sacha2018prl,Liang2018NJP}.
For $\lambda\ll 1$, the interaction is Coloumb-like \cite{guo2016pra,Liang2018NJP}, 
$
U(\tb{Z}_i-\tb{Z}_j)=\gamma \pi^{-1}|\tb{Z}_i-\tb{Z}_j|^{-1}.
$ Thus, we arrive at the many-body Hamiltonian 
\begin{eqnarray}\label{HU}
H=\sum_iH_{s}(\tb{Z}_i)+\frac{\gamma}{\pi }\sum_{i<j}\frac{1}{|\tb{Z}_i-\tb{Z}_j|}.
\end{eqnarray}
By introducing the {\it phase space force} $ \tb{F}_i\equiv -\nabla_iH$ with $\nabla_i\equiv(\partial/\partial X_i,\partial /\partial P_i)$ and  the unit direction vector $\hat{\tb n}$ perpendicular to the phase space plane, we can rewrite Hamilton's canonical equations as 
\bea\label{ZnF}
\frac{d}{dt}{\tb{Z}}_i=\hat{\tb n}\times {\tb{F}}_i.
\eea
As a result, the phase space force causes a displacement of the atoms perpendicular to the force direction in phase space, which is similar to the Lorentz force.

In the presence of dissipation, the stable points become attractors \cite{ott2002}. In other words, a non-interacting atom tends to relax towards the closest stable point. Introducing a sufficiently strong repulsive interaction  $(\gamma>0)$ and initially preparing the atoms close to the origin will, thus, give rise to  a disk-shaped crystal, see Fig.~\ref{Fig_Honeycomb}(c). 

It is important to keep in mind that in a finite geometry the atom-atom interactions  tend  to distort the equilibrium configuration. We now want  to discuss how to mitigate this undesired effect.
In a mean-field approximation, the effective electrostatic potential experienced by an atom at phase-space position $\mathbf{Z}_i$ (induced by  the remaining atoms)   can be approximated by the Coulomb potential for a uniformly charged disk  
\bea\label{eq-Udisc}
\overline{U}(\tb Z_i)=2\pi\sigma R\Big[1-\sum_{l=0}(2l+1)\Big[\frac{(2l-1)!!}{(2l+2)!!}\Big]^2\Big(\frac{|\tb Z_i|}{R}\Big)^{2l+2}\Big].\nl
\eea
Here, $\sigma=\gamma/\sqrt{3}\pi^3$ is the effective charge density and we have assumed that the disk of radius $R$ is centered about the origin of phase space.  We note that the leading order term ($l=0$) in the above effective potential represents a harmonic potential $\overline{U}(\tb Z_i)\sim-\sigma\pi R^{-1}|\tb Z_i|^2/2$. We can further improve the fitting of $\overline{U}(\tb Z_i)$ to a parabolic potential (and, thus, partially taking into account  the  high-order $l$-th terms) by adjusting the curvature 
$$
\overline{U}(|\tb Z_i|)\sim-1.35 \frac{ \sigma \pi}{2R}|\tb Z_i|^2.
$$
This effective parabolic potential  can be easily  counter-balanced by a laser-generated  potential, thereby, strongly reducing the lattice distortion. In practice, this can be achieved simply by choosing the appropriate detuning $\delta\omega=1.35\sigma\pi R^{-1}|\tb Z|^2/2$ in Hamiltonian (\ref{PSL}) instead of resonant driving, cf. Fig.~\ref{Fig-SM-Chiral}(a).

\subsection{Dynamics of crystal vibrations}
Each atom oscillates around its equilibrium position, i.e., $\tb{Z}_{L}(t)=\tb{Z}^0_{L}+{\tb u}_{L}(t)$ where   ${\tb u}_{L}=(u_{L}^X,u_{L}^P)$ is the displacement on the lattice site $L$. 
The equilibrium points are determined by 
${\partial H}/{\partial u_{L}^\alpha} =0$
($\alpha\in \{X,P\}$).
By expanding the Hamiltonian Eq.~(\ref{HU}) up to second order, we rewrite it in terms of the classical variable $\alpha_{L}\equiv \frac{1}{\sqrt{2\lambda}}\big(u^X_{L}+iu^P_{L}\big)$ (see detailed derivation in Appendix \ref{App:LH}) 
\bea\label{mathcalH}
\mathcal{H}&=&\sum_L\omega_L\alpha^*_{L}\alpha_{L}+\frac{1}{2}g_{L}\alpha^{*2}_{L}+\frac{1}{2}g_{L}^*\alpha^2_{L}\nl
&&-\frac{\gamma}{4\pi}\sum_{L\neq L'}\frac{\alpha^*_{L}\alpha_{L'}+3e^{i2\varphi_{LL'}}\alpha^*_{L}\alpha^*_{L'}}{|\tb{Z}^0_L-\tb{Z}^0_{L'}|^3}+h.c.
\eea
Here $\omega_{L}$ are the onsite quasienergies with $\omega_{L}-\omega_0\sim \mathcal{O}(\gamma)$, and $\varphi_{LL'}$ is the angular coordinate of  $\tb{Z}^0_L-\tb{Z}^0_{L'}$, cf. Fig.~\ref{Fig_Honeycomb}(d), defined via
\bea\label{eq:varphi}
\left\{\begin{array}{l}
X_L-X_{L'}=R_{LL'}\cos\varphi_{LL'},\\ P_L-P_{L'}=R_{LL'}\sin\varphi_{LL'}.
\end{array}\right.\ \ 
\eea  The Hamiltonian Eq.~(\ref{mathcalH}) describes  phonons propagating in our honeycomb phase-space crystal. It  is reminiscent of the tight-binding model for electrons in graphene, but with two qualitatively new features: (i) Our phase-space phonons can hop between  any two arbitrarily distant sites reflecting the long-range nature of the atom-atom interaction, as indicated by the blue arrow in Fig.~\ref{Fig_Honeycomb}(d); (ii) The excitation-number is not conserved, because a phonon pair can be  created or annihilated on any pair of sites. 

It is important to highlight an important qualitative feature of our model for the phase-space vibrations: \textit{the complex phases $\varphi_{LL'}$ of the anomalous pairing interaction have a geometrical interpretation as the angular coordinate of the link connecting  two sites,} cf. Eq.~(\ref{eq:varphi}) and Fig.~\ref{Fig_Honeycomb}(d). The same interpretation will still hold for any arbitrary phase-space single-particle  Hamiltonian $H_s(\tb{Z})$ and atom-atom interaction $U(\tb{Z}_i-\tb{Z}_j)$ (see Appendix \ref{App:LH} for a more detailed discussion) and, thus, is to be viewed as a general property of vibrations in phase space.   This is in stark contrast to the non-reciprocal phases of tight-binding Hamiltonians in 2D real space which can be tuned without displacing the lattice sites. Below, we show that this complex phases lead to a topological phase transition.

\section{Topological Band Structure}\label{TP}

\subsection{Connection to QAHE}

\begin{figure}
\centerline{\includegraphics[width=0.9\linewidth]{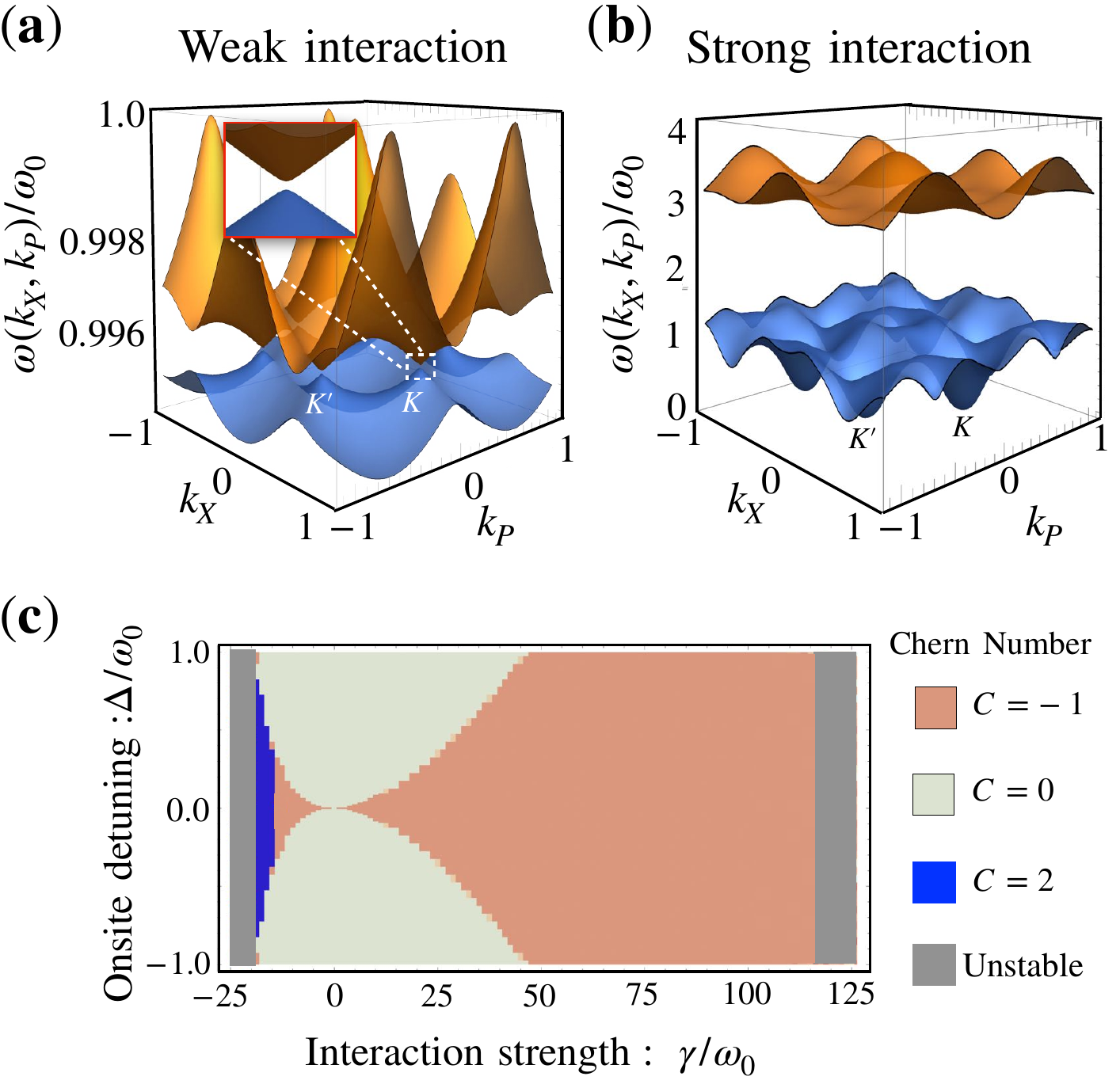}}
\caption{\label{Fig_Bandstructure}
 Bulk band structures for weak interaction strength $\gamma/\omega_0=-0.2$ {\bf (a)} and strong interaction $\gamma/\omega_0=102$ {\bf (b)}. Topological phase diagram {\bf (c)} with Chern number $C$ (the Chern number of the lowest band) as a function of interaction strength $\gamma$ and on-site detuning $\Delta$, which lifts the degeneracy at the $K$ ($K'$) symmetry points. }
\end{figure}

In order to better distinguish the effects of the (long-range) hopping and the pairing interaction, it is instructive to first consider a regime where the latter is suppressed. This will be the case if the creation of phonon pairs is far off-resonant, for $\gamma\ll\omega_0$. The band structure in this regime is shown in Fig.~\ref{Fig_Bandstructure}(a). While the long-range hopping strongly modifies it compared to its graphene counterpart, its main distinguishing feature is still on display: it supports two Dirac cones at the high-symmetry points $K$ and $K'$, with the tips of the cones separated by a small gap. We note that the Dirac cones are symmetry-protected (they should be gapless if the Hamiltonian has six-fold rotational symmetry and real hopping amplitudes). Thus, the small band gap must be induced by the complex amplitudes of the pairing perturbation. In 2D real space, the appearence of such complex amplitudes would be interpreted as being induced by the  breaking of the time-reversal symmetry and is well known to lead to a  topological  band gap \cite{Haldane1988PRL}.

This can be further substantiated by observing that, in the off-resonant regime, the main effect of the pairing interaction is to induce effective hopping transitions with non-reciprocal phases, cf. left subfigure in Fig.~\ref{Fig_Honeycomb}(e). Thus, a phonon propagating on a close loop $L(\cdot)$ will acquire a geometrical Berry phase $\Phi(L(\cdot))$.
These geometrical phases are similar to the Aharonov-Bohm phases acquired by charged particles in a magnetic field and are, thus, often referred to as synthetic magnetic fluxes. %
In fact, we can derive an effective particle-conserving description from a perturbation theory in $\gamma/\omega_0$.
That is to say, the pairing term $\propto \alpha^*_L\alpha^*_{L'}$ in the Hamiltonian~(\ref{mathcalH}) can be cancelled and replaced by a hopping term $\tilde{h}_{LL'}\alpha^*_L\alpha_{L'}$ with pairing-induced hopping rate given by (see the detailed derivation in Appendix~\ref{App:PISMF})
 \bea\label{}
&&\tilde{h}_{LL'}\approx-\Big(\frac{3\gamma}{2\pi }\Big)^2\sum_{\bar{L}}\frac{2}{\omega_{\bar{L}}}\frac{1}{R^6_{L\bar{L}}}e^{i2(\varphi_{L\bar{L}}-\varphi_{L'\bar{L}})}.
\eea 
In Fig.~\ref{Fig_Honeycomb}(e), we show the geometric angle $\tilde{\varphi}\equiv \varphi_{L\bar{L}}-\varphi_{L'\bar{L}}$ and the resultant phase exponent $2i\tilde{\varphi}$ of complex pairing-induced hopping rate.
This differs from the well-known Haldane model for the anomalous Quantum Hall effect \cite{Haldane1988PRL} only in that it includes long-range hopping transitions. As for the Haldane model, the  non-reciprocal hopping phases can be viewed as being induced by an effective staggered magnetic field, cf. right subfigure in Fig.1(e), and open a small topological gap of width $\sim \gamma^2/\omega_0$  and with band-gap Chern number $C=-1$, cf. Fig.~\ref{Fig_Bandstructure}(a). We emphasize that this synthetic \textit{staggered} magnetic field for phonons is a purely \textit{classical} effect   and is not to be confused with the \textit{uniform} synthetic magnetic field that can be introduced to describe the motion of a single \textit{quantum} atom  in a non-commutative phase space \cite{Guo2013PRL,Zhang2017pra,Liang2018NJP,Lorch_2019,Leboeuf1990PRL,Leboeuf1992Chaos,Liang2018NJP}.

\begin{figure*}
\centerline{\includegraphics[width=0.9\linewidth]{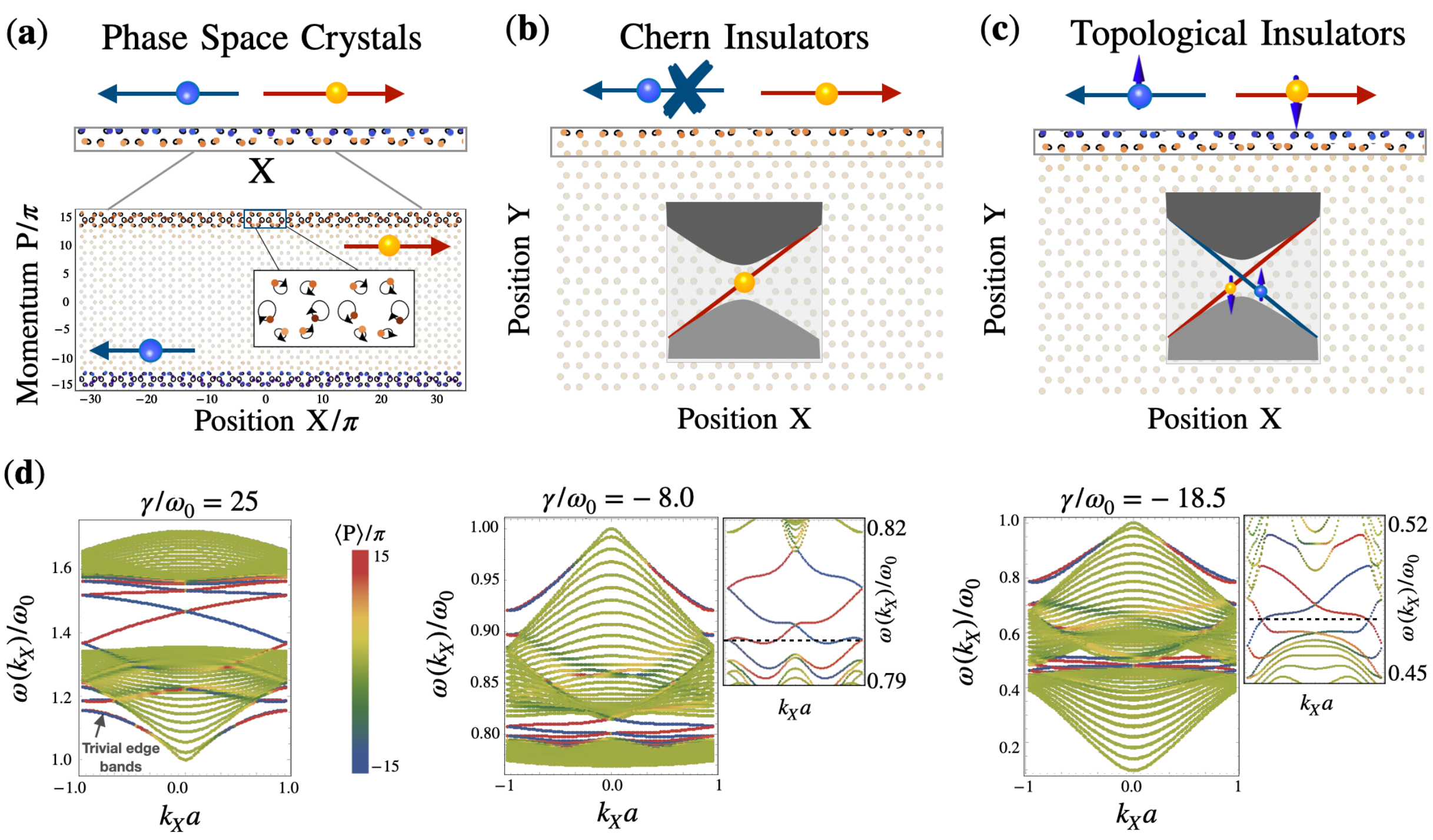}}
\caption{\label{Fig_TRE}
{\bf{Topological edge states.}}
{\bf (a)} Strip-shaped phase-space crystal with two topological edge states propagating on the boundaries. 
Inset shows atoms' orbits (circles) and vibration energy (colour). Upper figure represents 1D system in real space. 
{\bf (b)-(c)} Topological transport on the edge of 2D real-space Chern insulators (b), time-reversal symmetric topological insulators (c). The two inserts indicate typical topological band structures of the strip for each case.
{\bf (d)} Band structures of the strip-shaped phase space crystal for interaction strengths $\gamma/\omega_0=25$ (left),  $\gamma/\omega_0=-8.0$ (middle) and $\gamma/\omega_0=-18.5$ (right). The colour indicates the average $P$ of each eigenstate, the insets in the middle and right figures are the zoomed-in band structures around the energy gap and $a=4\pi/\sqrt{3}$ is the lattice constant in $X$ direction of the strip.
For  2D real-space Chern insulators (b), the edge states do not have a time-reversed partner leading to \textit{chiral} transport. For  time-reversal symmetric  topological insulators (c), each edge state has a time-symmetric partner co-localized on the same boundary but with opposite spin (or out-of-plane mirror symmetry) and propagation  direction. This leads  to \textit{helical} transport. Also for our phase-space crystal (a), an edge state  have a time-reversed partner (in the ideal case of two open separated boundaries). However, the two edge states (without any internal degree of freedom)  have the same chirality and are not co-localized in phase space.
}
\end{figure*}

\subsection{Strong interaction regime}
In the strong interaction regime, $|\gamma|\gg |\omega_0|$,  the phonon number is not even approximately conserved due to the pairing interactions in Hamiltonian (\ref{mathcalH}) and it is, thus, not possible to employ a single-particle description.  The physics is also different compared to superconducting systems \cite{Kotetes2013njp} (with fermionic pairing-interactions):  for bosons,  there is no limit to the occupation number of single-particle states, which can lead to the amplification of fluctuations and even to instabilities \cite{shindou2013prb,Peano2016NC,Bardyn_PRB_2016}.

In the stable regime, the Hamiltonian~(\ref{mathcalH}) can be diagonalized via a bosonic Bogoliubov transformation (see Appendix \ref{App:TBS}). For the bulk, $\mathcal{H}=\sum_{\tb k,n=1,2}\omega_{{\tb k},n}|\beta_{{\tb k},n}|^2$ where  $\tb k=(k_X,k_P)$ is the quasi-momentum, $\omega_{{\tb k},n}$ is the band structure, and $\beta_{{\tb k},n}$ are the normal modes,
 $$\beta^*_{\tb k,n}=\sum_{s=A,B}u_{n,s}(\tb k)\alpha^*_{\tb k,s}+v_{n,s}(\tb k)\alpha_{-\tb k,s}$$ with $\alpha_{\tb k,s}\equiv\frac{1}{\sqrt{N}}\sum_l\alpha_{l,s}e^{-i\tb k\cdot \tb Z^0_{l,s}}$ ($N$ is the number of primitive cells). The band Chern number is defined by  (see Appendix \ref{SCN})
\bea
C=\frac{1}{2\pi}\int_{BZ}\big(\nabla_k\times\mathcal{A}_1(\tb k)\big)\cdot\hat{z}
\eea
where
\begin{equation}\label{eq:Berry_conn_main}\mathcal{A}(\tb k)=i\sum_{s}[u^*_{1,s}(\mathbf{k})\nabla_{\tb k}u_{1,s}(\mathbf{k})-v^*_{1,s}(\mathbf{k})\nabla_{\tb k}v_{1,s}(\mathbf{k})]
\end{equation}
is the Berry connection for the lower band. Note that the definition of the Berry connection Eq.~(\ref{eq:Berry_conn_main}) has to be modified compared to the standard definition to account for the Bosonic nature of our excitations 
\cite{shindou2013prb}.
In a classical setting, the Chern number quantization as well as the other usual properties of the Chern number (including the bulk-boundary correspondence \cite{peano_topological_2018})  follow from the conservation of the Poisson brackets $\{\beta^*_{\tb k',n'},\beta_{\tb k,n}\}=i\delta_{n,n'}\delta_{\tb k',\tb k}$ (or, equivalently,
 $\sum_s |u_{n,s}|^2-|v_{n,s}|^2=1$).

The band structure for a comparatively strong interaction is shown in Fig.~\ref{Fig_Bandstructure}(b). In this case, the  Chern number $C=-1$ remains the same as in the weak-interaction limit. We note that close to the $K$ and $K'$-points (as labelled in the figure) the lowest band  approaches zero quasienergy. In the presence of  bosonic pairing interactions, the quasienergy can be viewed as the energy cost of producing a pair of Bogoliubov excitations. When this hits zero  for a critical threshold $\gamma^+_c\approx 115\omega_0$ or $\gamma^-_c\approx -20\omega_0$, the phase space crystal  becomes unstable, leading to a disordered gas phase.

\begin{figure*}
\center
\includegraphics[width=1\linewidth]{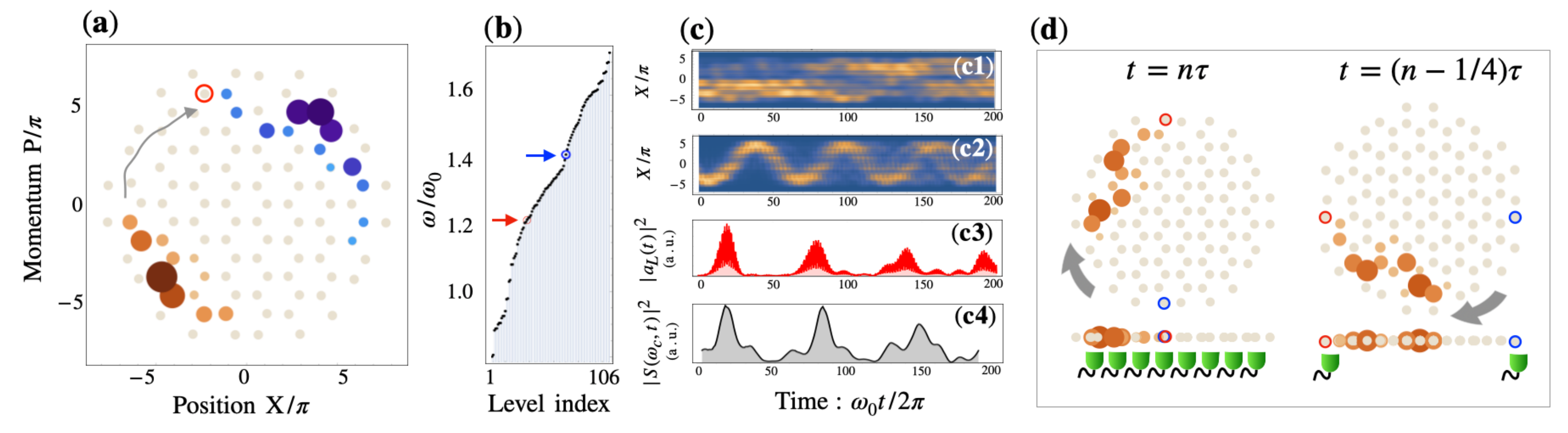}
\caption{{\bf{Chiral transport on the edge of disk-shaped phase-space crystal.}}
 {\bf (a)} One edge channel wave packet at different temporal instants. {\bf (b)} Energy spectrum with arrows indicating the energy level centres for constructing the wave packet in (a) by the Gaussian superposition of eigenstates around them. {\bf (c1)-(c2)} Spacetime plot of wave packet projected on the spatial $X$ dimension. The central frequencies $\omega_c$ of wave packets in (c1) and (c2) are indicated by the red and blue arrows in (d) respectively. {\bf (c3)} Vibration $|a_L(t)|^2$ of the atom marked by red circle in (a). {\bf (c4)} Finite-time-window power spectrum $S(\omega_c,t)$ of a collective signal from all the atoms. 
{\bf{(d)}} Detection of chiral direction. (Left) At stroboscopic time steps $t=n\tau$ with $n\in\mathbb{Z}$, an array of detectors are placed along the 1D system in real space to measure the distribution of vibrations over the 1D real space. (Right) At earlier time moments $t=(n-1/4)\tau$, the atoms (marked with red and blue circles) that are close in real space at stroboscopic time moment become separated at the two ends of 1D system in real space and thus can be detected independently.
[Parameters: interaction $\gamma=25\omega_0$ for all figures; disk radius $R=8\pi$ and $\delta\omega=1.35\sigma\pi R^{-1}$; time window $\Delta t=10\times 2\pi\omega^{-1}_0$ for (c4)]
}\label{Fig-SM-Chiral}
\end{figure*}

\subsection{Topological phase diagram}
 We have systematically investigated our phase space crystal by varying the interaction strength but also allowing for different onsite  quasienergies, $\omega_L\equiv\omega_0\pm2\Delta$, for the two honeycomb sublattices, cf. Eq.~(\ref{PSL}). The onsite detuning $\Delta$, which is engineered using additional stroboscopic lasers,      breaks the inversion  symmetry, allowing a trivial band gap. 
The ensuing topological phase diagram is shown in Fig.~\ref{Fig_Bandstructure}(c).  For $\Delta=0$, the phase-space crystal has Chern number $C=-1$ for a broad range of attractive and repulsive interactions, but can also switch to $C=2$ for negative interactions, before becoming unstable. The $C=2$ topological phase is not present in the  Haldane model  \cite{Haldane1988PRL}.  In the region of weak interactions, the band edge is cone-shaped and the band gap scales as  $\gamma^2/\omega_0$ for $\Delta=0$. In this regime, the gap determines how far the topological region extends into the $\Delta\neq 0$ region.

\section{Topological boundary states and time-reversal symmetry}\label{sec:TT}

In two-dimensional \textit{real} space, the time-reversal transformation changes the chirality of trajectories, e.g., from clockwise to anti-clockwise.  For this reason, robust chiral edge states -- without a time-reversed partner with opposite chirality -- can  be implemented in 2D real space only after breaking the TRS. However, this constraint  does not apply to phase-space crystals,  because  the chirality of motion in phase space remains unchanged under a time-reversal transformation. Indeed, because of the complex phases $\varphi_{LL'}$,  Hamiltonian (\ref{mathcalH}) does not support any local anti-unitary symmetry. This is, in fact, the standard formal precondition for non-trivial Chern numbers and chiral edge states \cite{Brouder_PRL_2007,Ryu_RevModPhys_2016}. For phase-space crystals, it can be fulfilled even though the time-reversal remains a symmetry, because this symmetry rearranges the phase-space crystal in a non-local fashion, i.e.,  $(X_i,P_i)\rightarrow(X_i,-P_i)$ for all the atoms.

It is interesting to investigate how this unusual status of the time-reversal symmetry bears on the topological edge states of a phase space crystal. For concreteness, we initially focus on the conceptually simple scenario of a strip. For simplicity we neglect the small  deformations of the lattice equilibrium configurations induced by the atom-atom interactions.

In Fig.~\ref{Fig_TRE}(a), we display two edge states for vibrations about an equilibrium configuration that is invariant under time-reversal symmetry (the amplitude of the vibrations is encoded using two different color scales to distinguish the two different states). In addition, we show the band structure for three different values of the system parameters in the topological phases $C=-1$ and $C=2$, cf. Fig.~\ref{Fig_TRE}(d). We note that for any energy inside the band gap the net number of topological edge states (difference between anti-clokwise movers and clockwise movers) on each of the boundaries  is equal to the band gap Chern number as predicted  by the bulk-boundary correspondence \cite{Rudner2013PRX}. According to this correspondence, the edge states on opposite phase-space boundaries have the same chirality and, thus, opposite propagation directions, cf. Fig.~\ref{Fig_TRE}(a). 

For the special case considered here, where the time-reversal is a symmetry,
the edge states on the two boundaries are obtained one from the other by applying the time-reversal transformation.  We note that  the two states are co-localized in real space, cf. upper figure in Fig.~\ref{Fig_TRE}(a). In spite of this, their coupling is exponentially suppressed with the phase space separation (in this geometry the width of the strip), leading to topologically robust transport. This is in stark contrast to  time-reversal symmetric topological insulators \cite{Kane2005PRL,Bernevig2006Science} whose time-reversal partner edge states  have opposite chirality, do not have any spatial separation and remain decoupled only as long as the time-reversal symmetry is not broken by the disorder, cf. Fig.~\ref{Fig_TRE}(c). Our situation is also different compared to standard  Chern insulators which do not support any  time-reversed partner solution, cf. Fig.~\ref{Fig_TRE}(b). This conclusion holds true also for Chern insulators  of systems with one real and one synthetic  dimension \cite{Celi2014PRL,Ozawa2019NRP}:  these systems will support counter-propagating edge states that are co-localized in the single  real dimension but  are NOT  time-reversal partners.

\section{Topological transport}\label{sec:TopTran}

We now turn to a realistic disk-shaped crystal with a randomly-shaped boundary [Fig.~\ref{Fig-SM-Chiral}(a)]. Here, we fully take into account the atom-atom interaction that deforms the equilibrium configuration. As discussed above, one can reduce the deformation using a finite  laser detuning $\delta\omega=1.35\sigma\pi R^{-1}|\tb Z|^2/2$ in Eq.~(\ref{PSL}), cf. discussion about Eq.~(\ref{eq-Udisc}). In Fig.~\ref{Fig-SM-Chiral}(a), we illustrate the chiral transport by tracking the time evolution of a wave packet with average quasienergy $\omega_c$ in the middle of the band gap, cf. blue arrow in Fig.~\ref{Fig-SM-Chiral}(b). It can be readily observed that the transport is robust against the boundary defects \cite{SuppVideo}. In Figs.~\ref{Fig-SM-Chiral}(c1)-(c2), we compare the non-chiral and chiral transports by projecting the  evolution of the wave packet onto the coordinate $X$ in the rotating frame.  The time evolution of a single atom's vibrational energy  likewise reveals the periodicity of the packet traversing the disk's circumference, cf. Fig.~\ref{Fig-SM-Chiral}(c3). However, in any real experiment, it might be easier to obtain a collective signal from all the atoms, for example by light scattering: $I(t)= \sum_j\cos[k_s X_j(t)]$ obtained by using a stroboscopic optical lattice for detection. One can extract the finite-time-window power spectrum  $$S(\omega,t)\propto \Big|\Big\langle (I(t)-\langle I\rangle)e^{i\omega t} \Big\rangle_{\Delta t}\Big|^2,$$ where $\big\langle \cdots \big\rangle_{\Delta t}$ represents the time average over a finite time window. In Fig.~\ref{Fig-SM-Chiral}(c4), we plot the power spectrum $S(\omega_c,t)$ (at $k_s=9.95$) as a function of time. The periodic temporal peaks of $S$ nicely indicate the chiral motion of the wave packet.

The parameters used for Figs.~\ref{Fig-SM-Chiral}(a)-(c) could be obtained in an experiment  with ultracold $^{87}Rb$ atoms. An appropriate experimental setup involves 8 lasers (4 for the static trapping, 2 for the stroboscopic trapping and 2  for the optomechanical detection), see sketch by Fig.~\ref{Fig_Honeycomb}(a). Assuming a realistic longitudinal trapping frequency $\omega_{ax}=2\pi\times \SI{50}{\hertz}$ and $\omega_0=0.1$ (to guarantee  slow vibrations in the rotating frame), the wavepacket's time of flight during a round-trip along the disk edge  would be approximately $2\pi\times 50/(\omega_{ax}\omega_0)\approx\SI{10}{\second}$, shorter than the typical lifetime of the atom cloud \cite{Davis1995prl,Cornell2002rmp,Greif2007thesis}. With $k^{-1}=\SI{45}{\micro\meter}$, one works in the semi-classical regime ($\lambda=0.04$). In the experiment, the transverse trapping frequency can reach up to $\omega_{tr}\approx 2\pi \times \SI{1.0}{\mega\hertz}$ \cite{Bloch2008RMP}. To avoid exciting the transverse mode during collisions, there is a restriction for the radius of the phase space crystal, $R<k\sqrt{\frac{\hbar\omega_{tr}}{m\omega^2_{ax}}}\approx 9.2\pi$, cf. Appendix~\ref{App:EP-SS}.
Taking into account the 3D scattering length $a=\SI{5.3}{\nano\meter}$ for  $^{87}Rb$ atoms, the 1D dimensionless interaction strength is $\gamma=2\hbar\omega_{tr} a k^3/(m\omega^2_{ax})\approx 1.25$, which can be further tuned by Feshbach resonance, cf. Appendix~\ref{App:EP-Int}, or adjusting $k$.

\subsection{Detection of chiral direction}

Measurement traces as in Figs.~\ref{Fig-SM-Chiral} (c2) (time trace of a quadrature) or (c3) (vibrational amplitude of a single-atom) would demonstrate the existence of robust chiral motion.  However, they do not yet allow to detect its chirality (clockwise or anti-clockwise).  
To infer the time trace of a quadrature  as displayed in Figs.~\ref{Fig-SM-Chiral}(c2), one could measure the local vibrations, as in Figs.~\ref{Fig-SM-Chiral}(c3), by placing detectors at the atom cloud, as illustrated by Fig.~\ref{Fig-SM-Chiral}(d). This seems to lead to an ambiguity: since the disk-shaped cloud is actually one-dimensional in real space it seems that any local measurement could not distinguish between atoms that are close in real space but have different momentum, e.g., the two atoms marked by the red and blue circles in the left panel of Fig.~\ref{Fig-SM-Chiral}(d). Without the ability to distinguish these opposite phase-space configuration, it is also  not possible to determine the direction of the chiral motion. However, it is important to keep in mind that the fixed equilibrium position of the atoms in the rotating frame should be viewed as the result of taking  a series of snapshots of the atom cloud at the stroboscopic time steps $t=n\tau$ with  $n\in \mathbb{Z}$ and $\tau$ being the stroboscopic time step. On the other hand, in the lab-frame continuous time picture, the whole phase space crystal structure rotates clockwise in phase space as indicated by the arrows in Fig.~\ref{Fig-SM-Chiral}(d). Importantly, the same experimental setup allows to monitor the motion for different stroboscopic time series,  e.g. with a one quarter of harmonic period earlier $t=(n-1/4)\tau$, $n\in \mathbb{Z}$. The resulting measurement trace allows to distinguish the atoms marked by the red and blue circles. Thus, by comparing the two traces taken with a different stroboscopic time series, one could deduce the chiral direction  from the delay between the vibrational peaks. This is even possible if the motion is only monitored close to the boundaries of the cloud (which should be experimentally easier), cf. the right panel of  Fig.~\ref{Fig-SM-Chiral}(d).


\begin{figure}
\centerline{\includegraphics[width=0.8\linewidth]{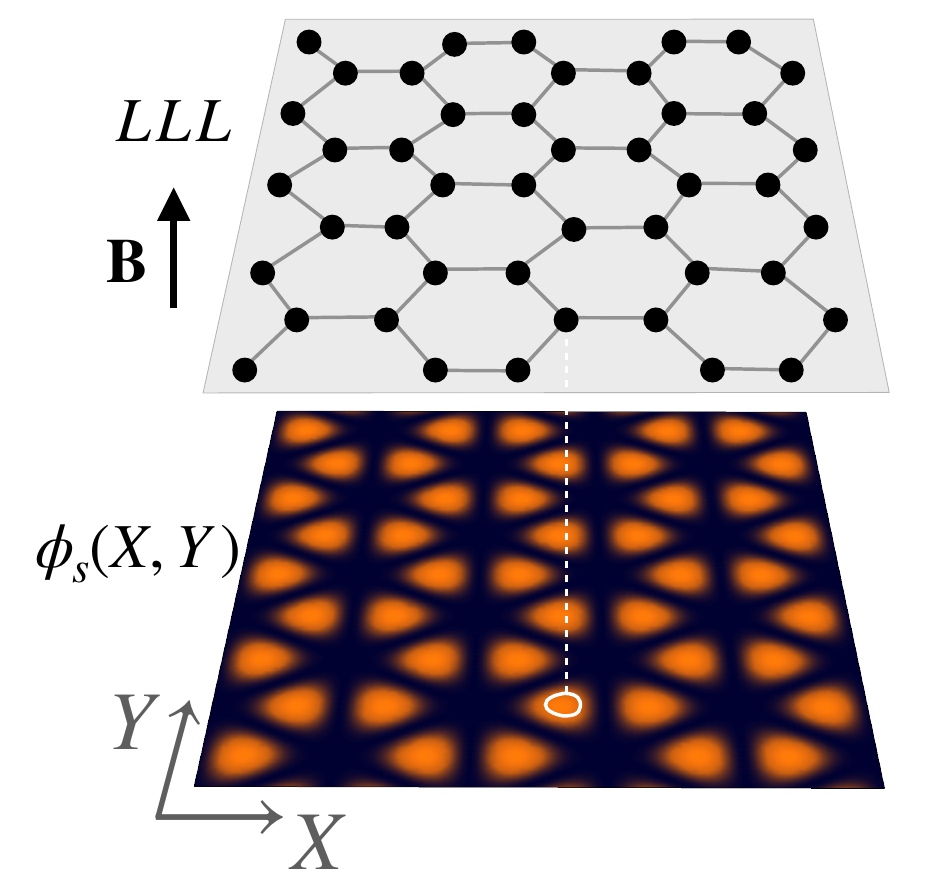}}
\caption{\label{Fig_LLL}{
 Real-space implementation for ions or semiconductor electrons - vibrations of cyclotron guiding centres in a strong magnetic field $\tb{B}$.
 Cyclotron orbital centres (upper panel) in LLL subjected to an external honeycomb lattice potential $\phi_s(X,Y)$ (lower panel).}
}
\end{figure}

\section{Implications for real space motion}\label{real-space-motion} 

The  phase-space dynamics investigated here could also be realized in 2D real space  using ions in a magnetic field. In an out-of-plane magnetic field $\tb{B}=B\hat{\tb n}$, the  motion of the ions can be decomposed into the cyclotron motion and the drift of the cyclotron guiding center. 
The two  guiding center coordinates $\mathbf{Z}=(X,Y)$ constitute an effective phase space due to $[X,Y]=-i\hbar/qB$ in quantum mechanics \cite{zak1997prl}. If all the relevant energy scales are much smaller than the cyclotron energy $\hbar\omega_c=\hbar|q|B/m$,  the motion is frozen into the lowest Landau level (LLL). The classical dynamics of the guiding center is governed by
\bea
\frac{d}{dt}\mathbf{Z}_i=(qB)^{-1}\tb{F}_i\times \hat{\tb n},
\eea
similar to the phase-space dynamics given by Eq.~(\ref{ZnF}). Considering that the phase-space interaction Eq.~(\ref{HU}) has the Coulomb form,  our model is implemented in the 2D external electrostatic potential $\phi_s(\tb{Z})\propto H_s(\tb{Z})$, cf. Eq.~(\ref{PSL}). We illustrate the real-space implementation for ions or semi-conductor electrons in Fig.~\ref{Fig_LLL}.

We note that for filling factors $\nu\equiv 2\pi n \hbar/eB$ (where $n$ is the density) below a critical threshold $\nu_w\sim \frac{1}{7}$, the ground state of the ions is expected to be a  crystal even without any external confining potential \cite{laughlin1983prl,tsui1988prl}, a so-called Wigner Crystal \cite{wigner1934pr,Jang2017np}. However, in this case the guiding centers are known to be arranged on a simple Bravais triangular lattice, leading to a single trivial phonon band \cite{bonsall1977prb,macdonald1990prl,Jang2017np}. Our work shows  that when the guiding centers are rearranged on a honeycomb lattice by an external electrostatic potential $\phi_s(X,Y)$, the crystal vibrations become topological. In this context, it is instructive to revisit our results in Fig.~\ref{Fig_Bandstructure}. The presence of topological phases with different sign of the Chern number $C$ indicates that the chirality of the vibrations can be reversed. Counterintuitively, for $C=2$ corresponding to strong repulsive interactions $\gamma$,  the chirality of the phonon edge states is opposite compared to the chirality of the cyclotron orbits, cf. Fig.~\ref{Fig-SM-Chiral}(d).

\section{Summary and Outlook}\label{outlook}

In summary, we have investigated the vibrational modes of phase space crystals formed by atoms that are periodically arranged in phase space.  We provided a general method for generating arbitrary lattice structures in phase space. We have shown that the small vibrations of this type of crystal structures can display gapless topologically robust chiral motion, even when the time-reversal is a symmetry. These topological phases are encoded in the symplectic Chern numbers of the bulk bands. Formally, the non-zero Chern numbers are induced by complex amplitudes of the two-mode pairing interactions. Interestingly, the complex phases of these amplitudes have a geometrical interpretation as the angular coordinate of the line connecting two atoms and can not be eliminated by a transformation that is local in phase-space. In addition, we have presented a realistic implementation  of our phase crystal and also discussed a scheme to detect the chiral vibrations. Finally, we explored the non-trivial implication of our investigation for the classical  motion of charged particles in 2D real space under a strong magnetic field and an additional electrostatic potential.

There are several possible research directions in the future beyond the present work. Robust topology produced by the combination of symplectic phase-space geometry and interactions represents a versatile concept that can be implemented in many physical platforms. For atoms with spin, one would obtain nonlocal spin-dependent interactions in phase space \cite{Liang2018NJP,Guo2020NJP}, coupling spin waves to topological phase-space phonons. Long-range real-space interactions permit the exploration of higher-dimensional generalizations of the physics discussed here in 1D. More complex driving can be used to synthesize arbitrary phase-space potentials \cite{guoinpre}, and single-shot measurements of multi-atom configurations will allow the observation of additional effects like nonlinear evolution.

\begin{acknowledgments}
 We acknowledge helpful discussions with Tirth Shah, Panagiotis Kotetes and Herrmann Schulz-Baldes.
\end{acknowledgments}

\appendix

\section{Arbitrary Phase Space Lattice}\label{App:PSLH}

\subsection{General form}
We discuss how to synthesize arbitrary lattice structures in phase space via multiple stroboscopic lattices. We start from the following generalised model of kicked harmonic oscillator 
\begin{equation}\label{KHO}
H_s=\frac12(x^2+p^2)+\sum_{n\in \mathbb{Z}}\sum_{q} K_q\cos (k_qx-\phi_q)\delta\Big(\frac{t}{\tau}-\theta_q-n\Big).
\end{equation}
Here $q$ represents the kicking sequence of stroboscopic lattices whose intensity $K_q$, wave vector $k_q$ and phase $\phi_q$ can be tuned at different time instances $t=\tau(n+\theta_q)$ with $n\in\mathbb{Z}$. To simplify the discussion, we first consider a single kick,
\begin{equation}
H_s=\frac12(x^2+p^2)+K_q\cos (k_qx-\phi_q)\delta\Big(\frac{t}{\tau}-\theta_q-n\Big).
\end{equation}
 We transform the above Hamiltonian into rotating frame with kicking frequency $2\pi/\tau$ using the generating function of the second kind
\begin{equation}\label{AppAG2}
G_2(x,P, t)=\frac{xP}{\cos (2\pi t/\tau)}-\frac{1}{2}x^2\tan \big(\frac{2\pi }{\tau}t\big)-\frac{1}{2}P^2\tan \big(\frac{2\pi }{\tau}t\big).
\end{equation}
The
corresponding canonical transformation is given by 
$$p=\frac{\partial G_2}{\partial x},\ \ \  X=\frac{\partial G_2}{\partial P},\ \ \ $$ which
results in the  transformation of phase-space coordinates
\bea\label{eq-SM-xpXP}
\left\{\begin{array}{l}
x=P\sin \big(\frac{2\pi t}{\tau}\big)+X\cos \big(\frac{2\pi t}{\tau}\big)
 \\
p=P\cos \big(\frac{2\pi t}{\tau}\big)-X\sin \big(\frac{2\pi t}{\tau}\big). 
\end{array}\right.\ \ 
\eea
and the transformed Hamiltonian
\begin{eqnarray}\label{HsXPt}
 &&H_s(X,P, t)=H_s(x,p, t)+\frac{\partial G_2}{\partial  t}=\frac{1}{2}\delta\omega(X^2+P^2)+\nl
&&K_q\cos \Big[k_q\big(P\sin \frac{2\pi t}{\tau}+X\cos \frac{2\pi t}{\tau}\big)-\phi_q\Big]\delta\Big(\frac{t}{\tau}-\theta_q-n\Big).\nl
\end{eqnarray}
Here, we have defined the global detuning parameter $\delta\omega\equiv1-2\pi/\tau$ between kicking and harmonic oscillation.
For weak resonant driving  ($|K_q|\ll 1$, $\delta\omega=0$), the single-particle dynamics can be separated by the fast harmonic oscillation and the low motion of its quadratures $(X,P)$.  The effective slow dynamics of quadratures $(X,P)$ is given by the lowest-order Magnus expansion, i.e., the time average of $H_s(X,P,t)$ in one kicking time period, 
\bea
H_s(X,P)&=&\frac{1}{\tau}\int_0^{\tau}H_s(X,P,t)dt\nl
&=&K_q\cos \Big[k_q\Big(P\sin 2\pi\theta_q+X\cos 2\pi\theta_q\Big)-\phi_q\Big].\nl
\eea
The above derivation is for classical dynamics. For quantum dynamics, the result has the same form with replacing $X$ and $P$ by their operators \cite{Billam2009PRA,Liang2018NJP}.
Extending the result for single kick to all kicks in Eq.~(\ref{KHO}), we obtain the general form of phase-space lattice Hamiltonian
\bea\label{HsXiPi}
H_s(X,P)
=\sum_{q}K_q\cos \Big[k_q\Big(P\sin 2\pi\theta_q+X\cos 2\pi\theta_q\Big)-\phi_q\Big].\nl
\eea
In principle, any arbitrary lattice Hamiltonian in phase space can be synthesized by multiple stroboscopic lattices.
%

\subsection{Honeycomb phase space lattice}\label{App:HPSL}
For the honeycomb lattice considered in our work, we can get the desired driving parameters by decomposing the honeycomb lattice into a series of cosine functions and comparing the series expansion to Eq.~(\ref{HsXiPi}). It is straightforward to check that the kicking sequence given by the first three lines of parameter table in Fig.~\ref{Fig-SM-kicking} 
generates the honeycomb lattice Hamiltonian in form of
\bea\label{SM-PSL}
H_s(\tb Z)&=&-\frac{16}{9}\omega_0\prod_{n=1}^3\sin^2\big(\frac{1}{2}\tb v_n\cdot\tb Z\big)-\frac{3}{2}\Lambda,\ \ \
\eea
where we have defined the   vector ${\tb Z}\equiv (X,P)$ and three ancillary vectors $\tb v_1=(\frac{2\sqrt{3}}{3},0)$, $\tb v_2=(-\frac{\sqrt{3}}{3},1)$, $\tb v_3=(-\frac{\sqrt{3}}{3},-1)$.
The extrema of
$H_s(X,P)$ represent stable points of the classical dynamics. The small vibrations about these  stable points have frequency $\omega_0=-9\Lambda$ from the linear expansion of Eq.~(\ref{SM-PSL}). The two sublattices of the phase-space honeycomb lattice Hamiltonian (\ref{SM-PSL}) have the same vibration frequency at their lattice sites.

One can also tune the on-site frequencies of two sublattices with the following phase space lattice Hamiltonian
\bea\label{onsitedetuning}
H^\Delta_s&=&-\frac{2\Delta}{\sqrt{3}}\sum_{n=1}^3\sin\big(\tb v_n\cdot\tb Z\big)\nl
&=&\frac{2\Delta}{\sqrt{3}}\sum_{n=1}^3\cos\big(\tb v_n\cdot\tb Z-\frac{3\pi}{2}\big),\ \ \ 
\eea
which can be implemented by additional stroboscopic lattices with kicking parameters $K_q={2\Delta}/{\sqrt{3}}$, $k_q={2}/{\sqrt{3}}$ and $\phi_q=3\pi/2$ for $\theta_q=1/12,5/12,9/12$ as listed by the fourth line of parameter table in Fig.~\ref{Fig-SM-kicking}. The on-site frequencies of two sublattices are then $\omega_0\pm 2\Delta$ with the on-site detuning paramter $\Delta$. 
The total single-particle Hamiltonian of honeycomb lattice is
\bea\label{Hs_SM}
H_s(\tb Z)&=&\frac{1}{2}\delta\omega|\tb Z|^2-\frac{16}{9}\omega_0\prod_{n=1}^3\sin^2\big(\frac{1}{2}\tb v_n\cdot\tb Z\big)\nl
&&-\frac{2\Delta}{\sqrt{3}}\sum_{n=1}^3\sin\big(\tb v_n\cdot\tb Z\big).
\eea
Here we have also included the global detuning $\delta\omega$ introduced in Eq.~(\ref{HsXPt}).

\section{Many-body Hamiltonian}

We consider many interacting particles trapped in the same 1D harmonic potential.
The interaction $V(x_i-x_j)$ induces an effective interaction of two particles on their quadratures \cite{sacha2015scirep,sacha2015pra,guo2016pra,Liang2018NJP}. The general method to extract this effective interaction in phase space has been developed in Refs.~\cite{guo2016pra,Liang2018NJP}. The interaction of cold atoms due to the $s$-wave scattering is a point-like contact interaction  $V(x_i-x_j)=\gamma\delta(x_i-x_j)$. The effective phase-space interaction potential is a Coulomb-like interaction
$
U(R_{ij})=\gamma \pi^{-1}R^{-1}_{ij},
$
where $$R_{ij}\equiv\sqrt{(X_i-X_j)^2+(P_i-P_j)^2}=\Big|\tb Z_i-\tb Z_j\Big|$$ is the phase-space distance between two atoms $i$ and $j$. The effective phase space interaction is valid for well-separated atoms which is the case of phase space crystals considered in this work \cite{guo2016pra,Liang2018NJP}. Finally, we have the many-body Hamiltonian for this work
\begin{eqnarray}\label{eq-Many-bodyH}
H&=&\sum_{L}H_s(X_{L},P_{L})+\frac{1}{2}\sum_{L\neq L'}U(R_{LL'})\nl
&\equiv& T+\Phi.
\end{eqnarray}
Here, we have relabelled the atoms by the subscript $L=(l,s)$ representing the $l$-th atom in the $s\in \{A,B\}$ sublattice. $T\equiv \sum_{L}H_s(X_{L},P_{L})$ is the total single-particle contribution and $\Phi\equiv \frac{1}{2}\sum_{L\neq L'}U(R_{LL'})$ othe total interaction part.

\subsection{Equilibrium configuration}\label{App:EC}
In the presence of the lattice potential and their effective interaction, the atoms have equilibrium configuration in phase space. The equilibrium points of atoms $\tb Z^0_L=(X^0_L,P^0_L)$ are determined by the condition
\bea
\frac{\partial H}{\partial X_L}\Bigg |_{(X^0_L,P^0_L)} =0,\ \ \ 
\frac{\partial H}{\partial P_L}\Bigg |_{(X^0_L,P^0_L)}  =0,
\eea
where $H$ is the many-body Hamiltonian (\ref{eq-Many-bodyH}). For the periodic boundaries in both $X$ and $P$ directions, the interactions from symmetric directions cancel each other and the equilibrium points are given by the  honeycomb lattice sites.
The periodic boundary is helpful for theoretical study. In the real experimental setup, however, the equilibrium positions of atoms will deviate from the lattice sites due to the open boundary of phase space crystal. In fact, the atoms tend to relax towards the stable points and concentrate about the origin of the phase space due to the presence of dissipation. By introducing a relative strong repulsive interaction  $(\gamma>0)$, the atoms will spread over the phase space and  form a disk-shaped crystal state  as shown by Fig.~1(c) in the main text. Note that  the interaction of the atoms on the disk will tend to distort the honeycomb equilibrium configuration. There is a mean-field potential induced by the interaction. The effective potential generated by the disk crystal in phase space can be approximated by the Coulomb potential on a uniformly charged disk plane given by Eq.(\ref{eq-Udisc}), and can be significantly counter-balanced by choosing the appropriate detuning $\delta\omega=1.35\sigma\pi R^{-1}|\tb Z|^2/2$ in the single-particle Hamiltonian (\ref{Hs_SM}) as discussed in Sec.~\ref{subsecMBD}.

\subsection{Linearised Hamiltonian}\label{App:LH}
We expand the total many-body Hamiltonian (\ref{eq-Many-bodyH}) around the equilibrium positions of atoms, i.e., $\tb{Z}_{L}(t)=\tb{Z}^0_{L}+{\tb u}_{L}(t)$ where   ${\tb u}_{L}=(u_{L}^X,u_{L}^P)$ is the displacement on the lattice site $L$. To the second order, the many-body Hamiltonian is given by (up to a constant)
\bea\label{etaH}
H=\frac{1}{2}\sum_{\alpha,\beta}\sum_{L,L'}\eta^{LL'}_{\alpha\beta}u_{L}^\alpha u_{L'}^\beta
\eea
with the matrix $\eta^{LL'}_{\alpha\beta}$ given by
\bea\label{psfm}
\eta^{LL'}_{\alpha\beta}=\frac{\partial^2H}{\partial u_L^\alpha\partial u_{L'}^\beta} \Bigg|_{\bf 0}=T_{\alpha\beta}^{LL'}+\Phi_{\alpha\beta}^{LL'}.
\eea
We call $\eta^{LL'}_{\alpha\beta}$ the \textit{phase space force matrix}, which means the resultant force along $\alpha$ direction exerted on the $L$-th atom due to the unit displacement along $\beta$ direction of the $L'$-th  atom.  
The contribution to $\eta^{LL'}_{\alpha\beta}$ from the single-particle Hamiltonian~(\ref{Hs_SM}) is
\bea\label{Talbe}
T_{\alpha\beta}^{LL'}&=&\frac{\partial^2H_s}{\partial u_L^\alpha\partial u_{L'}^\beta} \Bigg|_{\bf 0}=\frac{\partial^2H_s}{\partial u_L^\alpha\partial u_{L}^\beta} \Bigg|_{\bf 0}\delta_{LL'}\nl
&=&T^{LL}_{\alpha\beta}\delta_{LL'}=(\omega_0\pm 2\Delta)\delta_{\alpha\beta}\delta_{LL'}.
\eea
The contribution from interaction part can be obtained by calculating the  derivative of interaction potential 
\bea
\frac{\partial\Phi}{\partial Z_{L}^\alpha}&=&\sum_{L'\neq L}\frac{U'\big(R_{LL'}\big)}{R_{LL'}}\Big[(X_{L}-X_{L'})\delta_{X\alpha}+(P_{L}-P_{L'})\delta_{P\alpha}\Big].\nl
\eea
For two different atoms at different lattice sites ($L\neq L'$), we have the second derivative
\begin{widetext}
\bea\label{Phioff}
\Phi_{\alpha\beta}^{LL'}&=&
\frac{\partial^2\Phi}{\partial Z_{L}^\alpha\partial Z_{L'}^\beta }=
\frac{U'\big(R_{LL'}\big)}{R_{LL'}}\big(-\delta_{X\alpha}\delta_{X\beta}-\delta_{P\alpha}\delta_{P\beta}\big)+\frac{U''\big(R_{LL'}\big)R_{LL'}-U'\big(R_{LL'}\big)}{R^3_{LL'}}\nl
&&\times\Big[-(X_{L}-X_{L'})\delta_{X\beta}-(P_{L}-P_{L'})\delta_{P\beta}\Big]\Big[(X_{L}-X_{L'})\delta_{X\alpha}+(P_{L}-P_{L'})\delta_{P\alpha}\Big].\ \ \ \ \ \ 
\eea
\end{widetext}
For the atom at the single lattice site ($L= L'$), we have the second derivative
\begin{widetext}
\bea\label{PhiLL}
\Phi_{\alpha\beta}^{LL}&=&\frac{\partial^2\Phi}{ \partial Z_{ L}^\alpha\partial Z_{ L}^\beta}=\sum_{L'\neq L}\frac{U'\big(R_{ LL'}\big)}{R_{ LL'}}\big(\delta_{X\alpha}\delta_{X\beta}+\delta_{P\alpha}\delta_{P\beta}\big)+\frac{U''\big(R_{ LL'}\big)R_{ LL'}-U'\big(R_{ LL'}\big)}{R^3_{ LL'}}\nl
&&\times\big[(X_{ L}-X_{ L'})\delta_{X\beta}+(P_{ L}-P_{ L'})\delta_{P\beta}\big]\big[(X_{ L}-X_{ L'})\delta_{X\alpha}+(P_{ L}-P_{ L'})\delta_{P\alpha}\big]\nl
&=&\sum_{L'\neq L}(-1)\times\frac{\partial^2\Phi}{\partial Z_{ L'}^\beta \partial Z_{ L}^\alpha}.
\eea
\end{widetext}
We introduce the complex field at each lattice site 
$a_{L}(t)\equiv \frac{1}{\sqrt{2\lambda}}\big[u^X_{L}(t)+iu^P_{L}(t)\big],$
where $\lambda=\hbar k^2/m\omega_{ax}$ is the dimensionless Planck constant, and obtain the Hamiltonian from Eq.~(\ref{etaH}) 
\bea\label{SM-mathcalH}
\mathcal{H}/\lambda&=&\sum_{L,L'}h_{LL'}a^\dagger_{L}a_{L'}+\frac{1}{2}g_{LL'}a^\dagger_{L}a^\dagger_{L'}+\frac{1}{2}g_{L'L}^*a_{L}a_{L'}.\ \ \ \ \ \ 
\eea
 Here, $h_{LL}$ represents the onsite energy of one excitation on $L$-th lattice site,  $h_{LL'}(L\neq L')$ represents the hopping coefficient from $L'$-th lattice site to $L$-th lattice site, and $g_{LL'}$ represents pairing coefficient of creating or annihilating two phase-space phonons on the $L$-th and $L'$-th lattice sites. These coefficients are given by the phase space force matrix
\bea\label{hg}
\left\{\begin{array}{l}
h_{LL'}\equiv\frac{1}{2}\big(\eta_{XX}^{LL'}+\eta_{PP}^{LL'}\big)+i\frac{1}{2}\big(\eta_{PX}^{LL'}-\eta_{XP}^{LL'}\big)
 \\
g_{LL'}\equiv\frac{1}{2}\big(\eta_{XX}^{LL'}-\eta_{PP}^{LL'}\big)+i\frac{1}{2}\big(\eta_{PX}^{LL'}+\eta_{XP}^{LL'}\big). 
\end{array}\right.\ \ 
\eea
From the property $\eta_{\alpha\beta}^{LL'}=\eta_{\beta\alpha}^{L'L}$,
we have $h_{LL'}=h_{L'L}^*$ and $g_{LL'}=g_{L'L}$. 
Note that, compared to the Hamiltonian Eq.~(5) in the main text, we have replaced the variable $\alpha_L$ by $a_L=\alpha_L/\sqrt{\lambda}$ in the Hamiltonian Eq.~(\ref{SM-mathcalH}). In the linearised regime, these two descriptions are equivalent to each other, but the Hamiltonian Eq.~(\ref{SM-mathcalH}) here bears the advantage that it can be directly translated into the second-quantized description by taking $a_L$ as the ladder operator on the $L$-th lattice site, which makes the connection to other models, specifically the Haldane model \cite{Haldane1988PRL}, more transparent.

Below, we calculate the explicit form of $h_{LL'}$ and $g_{LL'}$ for the case of Coulomb-type interaction $U(R_{LL'})=\gamma \pi^{-1}R^{-1}_{LL'}$.
For the off-site ($L\neq L'$) coefficients, we have the hopping coefficients 
\bea
h_{LL'}&=&-\frac{U'\big(R_{ LL'}\big)}{R_{ LL'}}-\frac{1}{2}\frac{U''\big(R_{ LL'}\big)R_{ LL'}-U'\big(R_{ LL'}\big)}{R_{ LL'}}\nl
&=&-\frac{1}{2}\frac{U''\big(R_{ LL'}\big)R_{ LL'}+U'\big(R_{ LL'}\big)}{R_{ LL'}}\nl
&=&-\frac{\gamma}{2\pi R^3_{LL'}}
\eea
and the pairing coefficients
\bea
g_{LL'}&=&-\frac{1}{2}\frac{U''\big(R_{ LL'}\big)R_{ LL'}-U'\big(R_{ LL'}\big)}{R^3_{ LL'}}\nl
&&\times\Big[(X_L-X_{L'})^2-(P_L-P_{L'})^2\nl
&&+i2(X_L-X_{L'})(P_L-P_{L'})\Big]\nl
&=&-\frac{1}{2}\frac{U''\big(R_{ LL'}\big)R_{ LL'}-U'\big(R_{ LL'}\big)}{R_{ LL'}}e^{i2\varphi_{LL'}}\nl
&=&-\frac{3\gamma}{2\pi R^3_{LL'}}e^{i2\varphi_{LL'}},
\eea
where the phase parameter $\varphi_{LL'}$ is defined via
\bea
\left\{\begin{array}{l}
X_L-X_{L'}=R_{LL'}\cos\varphi_{LL'},\\ P_L-P_{L'}=R_{LL'}\sin\varphi_{LL'}.
\end{array}\right.\ \ 
\eea 
For the on-site ($L= L'$) coefficients, we have the on-site energy
\bea
\omega_L&\equiv& h_{LL}\nl
&=&\frac{1}{2}(T^{LL}_{XX}+T^{LL}_{PP})-\frac{1}{2}\sum_{L'\neq L}(\Phi^{LL'}_{XX}+\Phi^{LL'}_{PP})\nl
&=&\frac{1}{2}(T^{LL}_{XX}+T^{LL}_{PP})\nl
&&+\frac{1}{2}\sum_{L'\neq L}\frac{U''\big(R_{ LL'}\big)R_{ LL'}+U'\big(R_{ LL'}\big)}{R_{ LL'}}\nl
&=&\frac{1}{2}(T^{LL}_{XX}+T^{LL}_{PP})+\sum_{L'\neq L}\frac{\gamma}{2\pi R^3_{LL'}}.
\eea
and the squeezing rates 
\bea\label{gLLSM}
g_{L}&\equiv& g_{LL}\nl
&=&\frac{1}{2}(T^{LL}_{XX}-T^{LL}_{PP})+iT^{LL}_{PX}+\sum_{L'\neq L}\frac{3\gamma}{2\pi R^3_{LL'}}e^{i2\varphi_{LL'}}.\nl
\eea
The above expressions are valid for any boundary condition and arbitrary single-particle Hamiltonian. Finally, we obtain the Hamiltonian of phase-space lattice waves
\bea\label{HPSL}
\frac{\mathcal{H}}{\lambda}&=&\sum_L\omega_La^\dagger_{L}a_{L}+\frac{1}{2}g_{L}a^{\dagger2}_{L}+\frac{1}{2}g_{L}^*a^2_{L}\nl
&&-\frac{\gamma}{4\pi}\Bigg(\sum_{L\neq L'}\frac{a^\dagger_{L}a_{L'}+3e^{i2\varphi_{LL'}}a^\dagger_{L}a^\dagger_{L'}}{|\mathbf{Z}^0_L-\mathbf{Z}^0_{L'}|^3}+h.c.\Bigg). \ \ \ \ \ \ \ \ 
\eea
For the honeycomb single-particle Hamiltonian (\ref{Hs_SM}) and periodic boundary conditions, we have the simplified results of on-site coefficients
\bea\label{eq-app-omegaL}
\omega_{L}=\omega_0\pm 2\Delta+\sum_{L'\neq L}\frac{\gamma}{2\pi R^3_{LL'}}, \ \ \ \ g_{L}=0,
\eea
where the summation in Eq.~(\ref{gLLSM}) disappears due to the honeycomb lattice symmetry.
The Hamiltonian Eq.~(\ref{HPSL}) describes  phonons propagating in our honeycomb phase-space crystal. It  is reminiscent of the tight-binding model for electrons in graphene but our phase-space phonons can hop between  any two arbitrarily distant sites reflecting the long-range nature of the atom-atom interaction. 

\subsection{Pairing-induced staggered magnetic field}\label{App:PISMF}

 In order to better distinguish the effects of the (long-range) hopping and the pairing interaction, it is instructive to first consider a regime where the latter is suppressed, i.e., the creation of phonon pairs is far off-resonant  $|\gamma|\ll|\omega_0|$. In this case, we can cancel the pairing-interaction terms in the Hamiltonian (\ref{SM-mathcalH}) using the following unitary transformation
\bea
U\equiv \exp\Big(\sum_{\bar{L},\bar{L'}}\xi_{\bar{L}\bar{L'}}a^\dagger_{\bar{L}}a^\dagger_{\bar{L'}}
-\xi^*_{\bar{L'}\bar{L}}a_{\bar{L}}a_{\bar{L'}}\Big)\equiv e^{Y}.
\eea
It can be checked that $U^\dagger=e^{-Y}$ and thus $U^\dagger U=1$. Using the following identities
\bea
\big[a^\dagger_{ L}a_{ L'},a^\dagger_{\bar{L}}a^\dagger_{\bar{L'}}\big]&=&a^\dagger_{ L}a^\dagger_{\bar{L'}}\delta_{L'\bar{L}}+a^\dagger_{ L}a^\dagger_{\bar{L}}\delta_{L'\bar{L'}}\nl
\big[a^\dagger_{ L}a_{ L'},a_{\bar{L}}a_{\bar{L'}}\big]&=&-a_{\bar{L'}}a_{ L'}\delta_{L\bar{L}}-a_{\bar{L}}a_{ L'}\delta_{L\bar{L'}}\nl
\big[a^\dagger_{ L}a^\dagger_{ L'},a_{\bar{L}}a_{\bar{L'}}\big]&=&-a^\dagger_{ L}a_{\bar{L'}}\delta_{L'\bar{L}}-a^\dagger_{ L}a_{\bar{L}}\delta_{L'\bar{L'}}\nl
&&-a_{\bar{L'}}a^\dagger_{ L'}\delta_{L\bar{L}}-a_{\bar{L}}a^\dagger_{ L'}\delta_{L\bar{L'}}\nl
\big[a_{ L}a_{ L'},a^\dagger_{\bar{L}}a^\dagger_{\bar{L'}}\big]&=&a_{ L}a^\dagger_{\bar{L}}\delta_{L'\bar{L'}}+a_{ L}a^\dagger_{\bar{L'}}\delta_{L'\bar{L}}\nl
&&+a^\dagger_{\bar{L'}}a_{ L'}\delta_{L\bar{L}}+a^\dagger_{\bar{L}}a_{ L'}\delta_{L\bar{L'}},
\eea
we obtain
\bea
\big[a^\dagger_{ L}a_{ L'},Y \big]&=&\sum_{\bar{L}}\tilde{\xi}_{\bar{L}L'}a^\dagger_{ L}a^\dagger_{\bar{L'}}+\tilde{\xi}^*_{\bar{L}L}a_{\bar{L}}a_{ L'},\nl
\big[a^\dagger_{ L}a^\dagger_{ L'},Y \big]&=&\sum_{\bar{L}}\tilde{\xi}^*_{\bar{L}L'}a^\dagger_{ L}a_{\bar{L}}+\tilde{\xi}^*_{\bar{L}L}a_{\bar{L}}a^\dagger_{ L'},\nl
\big[a_{ L}a_{ L'},Y \big]&=&\sum_{\bar{L}}\tilde{\xi}_{\bar{L}L'}a_{ L}a^\dagger_{\bar{L}}+\tilde{\xi}_{\bar{L}L}a^\dagger_{\bar{L}}a_{ L'},
\eea
where we have defined
$
\tilde{\xi}_{\bar{L}L'}\equiv\xi_{\bar{L}L'}+\xi_{L'\bar{L}}
$
with $\tilde{\xi}_{\bar{L}L'}=\tilde{\xi}_{L'\bar{L}}$. The Hamiltonian transformed by $U$ to the leading order 
\bea
U^\dagger \mathcal{H}U/\lambda =\mathcal{H}/\lambda+\big[\mathcal{H}/\lambda,Y\big]+\cdots
\eea
where we have used the Hausdorff expansion $$e^{-Y}\mathcal{H}e^Y=\mathcal{H}+[\mathcal{H},Y]+\frac{1}{2!}[[\mathcal{H},Y],Y]+\cdots.$$ The leading-order correction is
\bea
\big[\mathcal{H}/\lambda,Y\big]&=&\sum_{L;L';\bar{L}}h_{LL'}\xi_{\bar{L}L'}a^\dagger_{ls}a^\dagger_{\bar{L'}}+h_{LL'}\xi^*_{\bar{L}L}a_{\bar{L}}a_{L'}\nl
&&+\frac{1}{2}g_{LL'}\tilde{\xi}^*_{\bar{L}L'}a^\dagger_{L}a_{\bar{L}} +\frac{1}{2}g_{LL'}\tilde{\xi}^*_{\bar{L}L}a_{\bar{L}}a^\dagger_{L'} \nl
&&+\frac{1}{2}g^*_{L'L}\tilde{\xi}_{\bar{L}L'}a_{L}a^\dagger_{\bar{L}} +\frac{1}{2}g^*_{L'L}\tilde{\xi}_{\bar{L}L}a^\dagger_{\bar{L}}a_{L'} \nl
&=&\sum_{L;L'}\tilde{h}_{LL'}a^\dagger_{L}a_{L'}+\frac{1}{2}\tilde{g}_{LL'}a^\dagger_{L}a^\dagger_{L'}+\frac{1}{2}\tilde{g}^*_{L'L}a_{L}a_{L'}.\nl
\eea
Neglecting constants from $a_La^\dagger_L=a^\dagger_L a_L+1$ and using the symmetry property $g_{LL'}=g_{L'L}$, we have the coefficients 
\bea
\tilde{h}_{LL'}&=&\sum_{\bar{L}}g^*_{\bar{L}L'}\tilde{\xi}_{\bar{L}L}+g_{L\bar{L}}\tilde{\xi}^*_{\bar{L}L'},\nl
\tilde{g}_{LL'}&=&\sum_{\bar{L}}h_{L\bar{L}}\tilde{\xi}_{\bar{L}L'},\nl
\tilde{g}^*_{L'L}&=&\sum_{\bar{L}}h_{\bar{L}L'}\tilde{\xi}^*_{\bar{L}L}.
\eea
To cancel the pairing terms in the Hamiltonian (\ref{SM-mathcalH}), it needs
$g_{LL'}+\tilde{g}_{LL'}=0$, i.e.,
\bea
g_{LL'}=-\sum_{\bar{L}}h_{L\bar{L}}\tilde{\xi}_{\bar{L}L'}.
\eea
We make the assumption that the on-site energy is much larger than the off-site hopping terms, i.e., $\omega_{L}\equiv h_{LL}\gg h_{L\bar{L}}$ with $L\neq\bar{L}$, which results in 
$$
\tilde{\xi}_{LL'}\approx -\frac{g_{LL'}}{\omega_{L}}\ \ \Longrightarrow\ \ 
\xi_{LL'}=\xi_{L'L}=-\frac{g_{LL'}}{2\omega_{L}}.
$$
Finally, we have the pairing-induced hopping rate
\bea\label{hLL}
\tilde{h}_{LL'}\approx-\sum_{\bar{L}}\frac{2}{\omega_{\bar{L}}}g_{L\bar{L}}g^*_{\bar{L}L'}.
\eea
Note that the pairing-induced hopping is proportional to the square of interaction strength $|\tilde{h}_{L\bar{L}}|\propto\gamma^2/\omega_0$.
Considering the case beyond NN interaction,
there is a direct real-valued hopping rate between lattice sites $L$ and $\bar{L}$, which is proportional to the interaction 
strength $h_{L\bar{L}}\propto\gamma$. As a result, the pairing-induced non-reciprocal hopping pathway $\tilde{h}_{L\bar{L}}$ interferes with a direct hopping pathway $h_{L\bar{L}}$ leads to a weak staggered magnetic flux  of the order $\Phi\sim\gamma/\omega_0$. This is the essential ingredients of Haldane model \cite{Haldane1988PRL}, which introduces a staggered magnetic field breaking the time-reversal symmetry inside each unit cell but leaving the total net flus through the unit cell zero. 

Our model recovers the anomalous quantum Hall effect similar to the Haldane model only in the weak off-resonant interaction regime. Our effective particle-conserving description (which is derived from a perturbation theory in $\gamma/\omega_0$) differs from the Haldane model in that it includes long-range hopping transitions. For sufficiently strong interaction (non-perturbative regime), our model has some new features which do not appear in Haldane model, e.g., new topological phase regime and the lost of stability.

\section{Topological Band Structure}\label{App:TBS}

\subsection{Periodic boundary condition}\label{PBC}

For periodic boundaries both in $X$ and $P$ directions, we Fourier transform the linearised Hamiltonian (\ref{SM-mathcalH}) using
\begin{eqnarray}
\left\{\begin{array}{l} a_{l,s}=\frac{1}{\sqrt{N}}\sum_{\tb k}a_{\tb k,s}\exp(i\tb k\cdot \tb Z^0_{l,s}); \\
  a^\dagger_{l,s}=\frac{1}{\sqrt{N}}\sum_{\tb k}a^\dagger_{\tb k,s}\exp(-i\tb k\cdot \tb Z^0_{l,s}),
\end{array}\right.
\end{eqnarray}
where $\tb k=(k_X,k_P)$, $Z^0_{l,s}$ is the equilibrium position of atom $L=(l,s)$ and $N$ is the total number of unite cells. The operators from inverse Fourier transformation
\begin{eqnarray}
\left\{\begin{array}{l} a_{\tb k,s}=\frac{1}{\sqrt{N}}\sum_{l}a_{l,s}\exp(-i\tb k\cdot \tb Z^0_{l,s}); \\
  a^\dagger_{\tb k,s}=\frac{1}{\sqrt{N}}\sum_{l}a^\dagger_{l,s}\exp(i\tb k\cdot \tb Z^0_{l,s}),
\end{array}\right.
\end{eqnarray}
satisfy the commutation of Bosons, i.e.,
\bea
\left.\begin{array}{cc}
[a_{\tb k,s},a^\dagger_{\tb k',s'}]=\delta_{\tb k \tb k'} \delta_{ss'};& 
[a_{\tb k,s},a_{\tb k',s'}]=0.
\end{array}\right.
\eea
The hopping terms in the Hamiltonian (\ref{SM-mathcalH}) become
\bea
&&\sum_{l,l'}h_{(l,s),(l',s')}a^\dagger_{l,s}a_{l',s'}\nl
&=&\frac{1}{N}\sum_{\tb k,\tb k'}a^\dagger_{\tb k,s}a_{\tb k',s'}\nl
&&\times\sum_{l,l'}h_{(l,s),(l',s')}e^{-i\tb k\cdot (\tb Z^0_{l,s}-\tb Z^0_{l',s'})-i(\tb k- \tb k')\cdot \tb Z^0_{l',s'}}\nonumber\\
&\equiv&\sum_{\tb k}h_{s,s'}(\tb k)a^\dagger_{\tb k,s}a_{\tb k,s'}
\eea
with the definition of hopping coefficients in reciprocal $\tb k=(k_X,k_P)$ space
\bea
h_{s,s'}(\tb k)\equiv\sum_{l}h_{(l,s),(l',s')}\exp[-i\tb k\cdot (\tb Z^0_{l,s}-\tb Z^0_{l',s'})].\ \ \ \ \ \ 
\eea
We find the property $h^*_{s,s'}( \tb k)=h_{s',s}( \tb k)$ using the identity $h_{(l,s),(l',s')}=h_{(l',s'),(l,s)}^*$ and the discrete translational lattice symmetry.
Similarly, the pairing terms in the Hamiltonian (\ref{SM-mathcalH}) become
\bea
&&\sum_{l,l'}g_{(l,s),(l',s')}a^\dagger_{l,s}a^\dagger_{l',s'}\nl
&=&\frac{1}{N}\sum_{\tb k,\tb k'}a^\dagger_{\tb k,s}a^\dagger_{\tb k',s'}\nl
&&\times\sum_{l,l'}g_{(l,s),(l',s')}e^{-i\tb k\cdot (\tb Z^0_{l,s}-\tb Z^0_{l',s'})-i(\tb k+\tb k')\cdot \tb Z^0_{l',s'}}\nonumber\\
&=&\sum_{k}g_{s,s'}(\tb k)a^\dagger_{k,s}a^\dagger_{-k,s'}
\eea
with the definition of pairing coefficient in reciprocal $\tb k=(k_X,k_P)$ space
\bea
g_{s,s'}(\tb k)\equiv\sum_{l}g_{(l,s),(l',s')}\exp[-i\tb k\cdot (\tb Z^0_{l,s}-\tb Z^0_{l',s'})].\ \ \ \ \ \ 
\eea
We find the property $g_{s,s'}(-\tb k)=g_{s',s}(\tb k)$ using the identity $g_{(l,s),(l',s')}=g_{(l',s'),(l,s)}$ and the discrete lattice translational symmetry.
As a result, the Hamiltonian in the reciprocal space is given by
\bea\label{Hkcom}
\mathcal{H}/\lambda&=&\sum_{\tb k,s,s'}h_{s,s'}(\tb k)a^\dagger_{\tb k,s}a_{\tb k,s'}\nl
&&+\frac{1}{2}g_{s,s'}(\tb k)a^\dagger_{\tb k,s}a^\dagger_{-\tb k,s'}+\frac{1}{2}g_{s,s'}^*(\tb k)a_{-\tb k,s'}a_{\tb k,s}\nonumber\\
&=&\sum_{\tb k} \tb A^T_{\tb k}\mathcal{H}_{\tb k}\tb A_{\tb k},
\eea
where we have deined the vector $\tb A_{\tb k}\equiv(a_{\tb k,A} , \  a_{\tb k,B},\   a^\dagger_{-\tb k,A}, \  a^\dagger _{-\tb k,B})^T$ and the Bogoliubov-de-Gennes (BdG) Hamiltonian
\bea
\mathcal{H}_{\tb k}&\equiv&
\left(\begin{array}{cccc} 
h_{A,A}(\tb k)  & h_{A,B}(\tb k) & \frac{1}{2}g_{A,A}(\tb k)  & \frac{1}{2}g_{A,B}(\tb k)  \\  
h_{B,A}(\tb k)  &  h_{B,B}(\tb k) & \frac{1}{2}g_{B,A}(\tb k)  & \frac{1}{2}g_{B,B}(\tb k)  \\  
\frac{1}{2}g_{A,A}^*(\tb k) & \frac{1}{2}g_{A,B}^*(\tb k)& h_{A,A}(-\tb k)  & h_{A,B}(-\tb k) \\
\frac{1}{2}g_{B,A}^*(\tb k)  & \frac{1}{2}g_{B,B}^*(\tb k)& h_{B,A}(-\tb k) & h_{B,B}(-\tb k) \nn
\end{array}\right).
\eea
From the Heisenberg equation 
\bea
-i\frac{d}{dt}a_{\tb k,s}&=&\frac{1}{\lambda}[\mathcal{H},a_{\tb k,s}]\nl
&=&\sum_{s'}-h_{s,s'}(\tb k)a_{\tb k,s'}\nl
&&-\frac{1}{2}g_{s,s'}(\tb k)a^\dagger_{-\tb k,s'}-\frac{1}{2}g_{s',s}(-\tb k)a^\dagger_{-\tb k,s'}\nonumber,\nonumber
\eea
we have the EOM for the ladder operator $\tb A_{\tb k}$ as following
\bea\label{EOMkxkp}
&&i\frac{d}{dt}\tb A_{\tb k}=
D(\tb k)
\tb A_{\tb k}.
\eea
Here $D(\tb k)$ is the dynamical matrix defined by
\bea
&&D(\tb k)\equiv\nl
&&\left(\begin{array}{cccc} 
h_{A,A}(\tb k)  & h_{A,B}(\tb k) & \bar{g}_{A,A}(\tb k) & \bar{g}_{A,B}(\tb k)   \\ 
h_{B,A}(\tb k)  & h_{B,B}(\tb k) & \bar{g}_{B,A}(\tb k) & \bar{g}_{B,B}(\tb k)  \\  
-\bar{g}^*_{A,A}(\tb k)  & -\bar{g}^*_{B,A}(\tb k) & -h^*_{A,A}(-\tb k) 
& -h^*_{A,B}(-\tb k) \\
-\bar{g}^*_{A,B}(\tb k)  & -\bar{g}^*_{B,B}(\tb k)& -h^*_{B,A}(-\tb k)
& -h^*_{B,B}(-\tb k) 
\end{array}\right)\nn
\eea
wit the symmetric pairing coefficients dedined by
$$
\bar{g}_{s,s'}(\tb k)\equiv\frac{1}{2}\Big[g_{s,s'}(\tb k)+g_{s',s}(-\tb k)\Big]. 
$$

The eigen solutions of EOM (\ref{EOMkxkp}) can be obtained by diagonalising the dynamical matrix $D(\tb k)$. We label the frequency spectrum by $\omega_{\tb k,n}$, where $n=1,2,3,4$ is the index of four bands. For each given $\tb k$ and band index $n$, the eigenstate is a four-component vector $|\tb{k},n\rangle=(u_{n,A},\ u_{n,B},\ v_{n,A},\ v_{n,B})$. Then, the Hamiltonian can be cast into a diagonal form of
\bea
\mathcal{H}/\lambda=\sum_{\tb k,n}\omega_{\tb k,n}b^\dagger_{\tb k,n}b_{\tb k,n},
\eea
where the normal modes are given by a Bogoliubov transformation in the form of $$b^\dagger_{\tb k,n}=\sum_{s=A,B}u_{n,s}(\tb k)a^\dagger_{\tb k,s}+v_{n,s}(\tb k)a_{-\tb k,s}.$$
From the property $\bar{g}_{s,s'}(\tb k)=\bar{g}_{s,s'}(-\tb k)$, the dynamical matrix has \textit{particle-hole symmetry} expressed by
\bea
\Xi D(\tb k) \Xi^{-1}=-D(-\tb k).
\eea
The particle-hole operator is defined via $\Xi=\tau_xK$ satisfies $\Xi^2=+1$, where 
\bea\label{taux}
\tau_x=\left(\begin{array}{cccc} 
0  & 0 & 1 & 0   \\ 
0  & 0 & 0 & 1  \\  
1  & 0 & 0 & 0  \\  
0  & 1 & 0 & 0  
\end{array}\right)
\eea
and $K$ is the complex conjugation. It follows that each eigenmode at frequency $\omega(\tb k)$ has a partner eigenmode at $-\omega(-\tb k)$, namely, creating a quasiparticle in the state $\omega(\tb k)$ has the same effect as removing one (creating a hole) from the state $-\omega(-\tb k)$. Therefore, we label the two upper bands ($n=1,2$) as particle bands and two lower bands ($n=3,4$) as the hole bands. From the bosonic commutation relationship $[b_{\tb k',n'},b^\dagger_{\tb k,n}]=\delta_{n,n'}\delta_{\tb k',\tb k}$, the eigenmodes have to follow  the   ortho-normal condition 
\bea\label{eq-sm-norm}
\langle {\tb k},n'|\Sigma_z|{\tb k},n\rangle\equiv\sum_s  u^*_{n',s}u_{n,s}-v^*_{n',s}v_{n,s}=\pm \delta_{n,n'},\nl
\eea
where $ \Sigma_z\equiv\mathrm{diag}(1,1,-1,-1)$ is a $4\times 4$ diagonal matrix and the positive (negative) sign corresponds to particle (hole) bands.

\subsection{Symplectic Chern number}\label{SCN}

Different from the particle-conserving case, the ground state of the BdG Hamiltonian $\mathcal{H}_{\tb k}$ is a multi-mode squeezed state with non-zero phonon/photon number. By regarding the $\tb k=(k_X,k_P)$ as an external adiabatic  parameter, we calculate the Berry phase accumulated by a single Bogoliubov quasi-particle in a specific $n$-th band along a closed loop covering the whole Brillouin zone (BZ), i.e.,
\begin{widetext}
\bea
\Phi_n&=&i\oint_{BZ} \langle S_{\tb k}|b_{\tb k,n}\nabla_{\tb k}b^\dagger_{\tb k,n}|S_{\tb k}\rangle\cdot d\tb k\nl 
&=&i\oint_{BZ} \langle S_{\tb k}|[b_{\tb k,n}\nabla_{\tb k}b^\dagger_{\tb k,n}]|S_{\tb k}\rangle\cdot d\tb k+i\oint_{BZ} \langle S_{\tb k}|\nabla_{\tb k}|S_{\tb k}\rangle\cdot d\tb k\nl 
&=&i\oint_{BZ} \Big[\sum_s  u^*_{n,s}(\tb k)\nabla_{\tb k}u_{n,s}(\tb k)-v^*_{n,s}(\tb k)\nabla_{\tb k}v_{n,s}(\tb k)\Big]\cdot d\tb k
+i\oint_{BZ} \langle S_{\tb k}|\nabla_{\tb k}|S_{\tb k}\rangle\cdot d\tb k\nl 
&\equiv&\oint_{BZ} \mathcal{A}_n(\tb k)\cdot d\tb k +\phi_n.
\eea
\end{widetext}
Here, $|S_{\tb k}\rangle$ is the ground state (Bogoliubov vacuum state) of $\mathcal{H}_{\tb k}$ and, in the second line, we have used $b_{\tb k,n}|S_{\tb k}\rangle=0$ by the definition of vacuum state. Because of the unusual ortho-normalization (\ref{eq-sm-norm}), we have updated the definition of the Berry connection by \cite{shindou2013prb,Peano2016NC}
\bea\label{eq-c5-An}
\mathcal{A}_n(\tb k)=i\langle \tb k,  n|\Sigma_z\nabla_{\tb k}|\tb k, n\rangle.
\eea
Note that the Berry connection defined here for the Bosonic many-body second-quantized Hamiltonian is different from the Berry connection for the single-particle Hamiltonian. 
The Bogoliubov vacuum $|S_{\tb k}\rangle$ depends on $\tb k$ and could possibly accumulate a Berry phase, i.e., $\phi_n\equiv i\oint \langle S_{\tb k}|\nabla_{\tb k}|S_{\tb k}\rangle\cdot d\tb k\neq 0$. 
However, the Berry phase of our interest is the additional Berry phase accumulated by the  quasi-particle, i.e., the difference of Bogoliubov vacuum Berry phase and that associated with a single quasiparticle excitation. We thus update the definition of Chern number by \cite{shindou2013prb,Peano2016NC}
\bea\label{eq-c5-Cn}
C_n=\frac{1}{2\pi}\int_{BZ}\big(\nabla_{\tb k}\times\mathcal{A}_n(\tb k)\big)\cdot\hat{\tb n}\in \mathbb{Z},
\eea
where $\hat{\tb n}$ is the unit vector normal to the $\tb k$-plane. Due to the additional element $\Sigma_{z}$ in the Berry connection (\ref{eq-c5-An}), the quantity given by Eq.~(\ref{eq-c5-Cn}) is also called \textit{symplectic Chern number}. Since the quantum states come back to itself along a closed loop, the quantities $\Phi_n$ and $\phi_n$ are integers multiple of $2\pi$, and thus the symplectic Chern number is also an integer.

As there is no net geometric phase (total flux of synthetic magnetic field) in our model, the sum of the Chern numbers over the particle bands must be zero. The Chern number of individual band may change after a phase transition where two or more bands touch each other, but their sum does not change. For this reason, we define the  Chern number of the lowest band as the Chern number of our system $C=C_1$.

\subsection{Strip boundary condition}

For the periodic boundary in $X$ direction but open boundary in $P$ direction (strip boundary condition), we can only perform Fourier transformation in $X$ direction
\begin{eqnarray}
\left\{\begin{array}{l} 
a_{l,s}=\frac{1}{\sqrt{N_X}}\sum_{k_X}a_{k_X,(l_P,s)}\exp(i k_X X^0_{l,s}); \\ 
a_{l,s}^\dagger=\frac{1}{\sqrt{N_X}}\sum_{k_X}a^\dagger_{k_X,(l_P,s)}\exp(-i k_X X^0_{l,s}),
\end{array}\right.
\end{eqnarray}
where $N_X$ is the number of unit cells in $X$ direction and $(l_P,s)$ labels the position of atoms in $P$ direction. Then, the hopping terms in the Hamiltonian (\ref{SM-mathcalH}) become
\bea
&&\sum_{l,l'}h_{(l,s),(l',s')}a^\dagger_{l,s}a_{l',s'}\nl
&=&\frac{1}{N_X}\sum_{l,l'}\sum_{k_X,k'_X}a^\dagger_{k_X,(l_P,s)}a_{k'_X,(l'_P,s')}\nl
&&\times h_{(l,s),(l',s')}e^{-i k_X(X^0_{l,s}-X^0_{l',s'})-i(k_X- k'_X)X^0_{l',s'}}\nonumber\\
&=&\sum_{l_P,l'_P}\sum_{k_X}h_{(l_P,s),(l'_P,s')}( k_X)a^\dagger_{k_X,(l_P,s)}a_{k_X,(l'_P,s')}\ \ \ \ \ \ 
\eea
with the Fourier transformation of hopping coefficients 
\bea
h_{(l_P,s),(l'_P,s')}( k_X)\equiv\sum_{l_X}h_{(l,s),(l',s')}e^{-i k_X\cdot (X^0_{l,s}-X^0_{l',s'})},\nl
\eea
which has the property $h^*_{(l'_P,s'),(l_P,s)}( k_X)=h_{(l_P,s),(l'_P,s')}( k_X)$ using the identity $h_{(l,s),(l',s')}=h_{(l',s'),(l,s)}^*$ 
and discrete translational symmetry in $X$-direction. Similarly, the pairing terms in the Hamiltonian (\ref{SM-mathcalH}) become
\bea
&&\sum_{l,l'}g_{s,s'}(l-l')a^\dagger_{l,s}a^\dagger_{l',s'}\nl
&=&\frac{1}{N_X}\sum_{l,l'}\sum_{k_X,k'_X}a^\dagger_{k_X,(l_P,s)}a^\dagger_{k'_X,(l'_P,s')}\nl
&&\times g_{(l,s),(l',s')}e^{-i k_X (X^0_{l,s}-X^0_{l',s'})-i( k_X+ k'_X)X^0_{l',s'}}\nonumber\\
&=&\sum_{l_P,l'_P}\sum_{k_X}g_{(l_P,s),(l'_P,s')}( k_X)a^\dagger_{k_X,(l_P,s)}a^\dagger_{-k_X,(l'_P,s')}\nl
\eea
with the Fourier transformation of pairing coefficients
\bea
g_{(l_P,s),(l'_P,s')}( k_X)\equiv\sum_{l_X}g_{(l,s),(l',s')}(l-l')e^{-i k_X\cdot (X^0_{l,s}-X^0_{l',s'})},\nn
\eea
 which has the property of $g_{(l_P,s),(l'_P,s')}(- k_X)=g_{(l'_P,s'),(l_P,s)}(k_X)$ using the identity $g_{(l,s),(l',s')}=g_{(l',s'),(l,s)}$
and discrete translational symmetry in $X$-direction.
Therefore, the Hamiltonian in momentum space is 
\bea\label{Hkcom1}
\mathcal{H}/\lambda&=&
\sum_{l_P,l'_P}\sum_{k_X}h_{(l_P,s),(l'_P,s')}( k_X)a^\dagger_{k_X,(l_P,s)}a_{k_X,(l'_P,s')}\nl
&&+\frac{1}{2}g_{(l_P,s),(l'_P,s')}( k_X)a^\dagger_{k_X,(l_P,s)}a^\dagger_{-k_X,(l'_P,s')}\nl
&&+\frac{1}{2}g^*_{(l_P,s),(l'_P,s')}( k_X)a_{k_X,(l_P,s)}a_{-k_X,(l'_P,s')}\nn
\eea
From the Heisenberg equation
\bea
&&-i\frac{d}{dt}a_{k_X,(l_P,s)}
=\sum_{l'_P,s'}-h_{(l_P,s),(l'_P,s')}( k_X)a_{k_X,(l'_P,s')}\nl
&&-\frac{1}{2}[g_{(l_P,s),(l'_P,s')}( k_X)+g_{(l'_P,s'),(l_P,s)}( -k_X)]a^\dagger_{-k_X,(l'_P,s')},\nl
\eea
we have EOM for the operator $\tb A_{l_p}(k_X)\equiv (a_{k_X,(l_P,A)},a_{k_X,(l_P,B)},a^\dagger_{-k_X,(l_P,A)},a^\dagger_{-k_X,(l_P,B)})^T$
\begin{widetext}
\bea
&&i\frac{d}{dt}\left(\begin{array}{c}  
\tb A_{1} \\ 
\vdots\\
\tb A_{l'_P} \\ 
\vdots \\  
\tb A_{N_P}  \end{array}
\right)=
\left(\begin{array}{ccccc} 
 D_{(l_P=1,l'_P=1)} & \cdots & D_{(l_P=1,l'_P)}&\cdots &  D_{(l_P=1,l'_P=N_P)}  \\  
\vdots & \ddots &\vdots  & \vdots & \vdots  \\ 
 D_{(l_P,l'_P=1)} & \cdots &  D_{l'_P,l_P} & \cdots &  D_{(l_P,l'_P=N_P)}  \\  
\vdots &\vdots& \vdots &\ddots  & \vdots \\
 D_{(l_P=N_P,l'_P=1)}  &\cdots &  D_{(l_P=N_P,l'_P)} &\cdots&  D_{(l_P=N_P,l'_P=N_P)}  
\end{array}\right)
\left(\begin{array}{c}  
\tb A_{1} \\ 
\vdots\\
\tb A_{l'_P} \\ 
\vdots \\  
\tb A_{N_P}  \end{array}
\right),\nonumber
\eea
\end{widetext}
where the block matrix element is given by 
\begin{widetext}
\bea
&&D_{(l_P,l'_P)}(k_X)\equiv\nl
&&
\left(\begin{array}{cccc} 
h_{(l_P,A),(l'_P,A)}(k_X) & h_{(l_P,A),(l'_P,B)}(k_X) & \bar{g}_{(l_P,A),(l'_P,A)}(k_X)  &\bar{g}_{(l_P,A),(l'_P,B)}(k_X)  \\  

h_{(l_P,B),(l'_P,A)}(k_X)  & h_{(l_P,B),(l'_P,B)}(k_X) & \bar{g}_{(l_P,B),(l'_P,A)}(k_X) & \bar{g}_{(l_P,B),(l'_P,B)}(k_X)  \\

-\bar{g}^*_{(l_P,A),(l'_P,A)}(-k_X)  & -\bar{g}^*_{(l_P,A),(l'_P,B)}(-k_X) & -h^*_{(l_P,A),(l'_P,A)}(-k_X) & -h^*_{(l_P,A),(l'_P,B)}(-k_X) \\

-\bar{g}^*_{(l_P,B),(l'_P,A)}(-k_X)  &-\bar{g}^*_{(l_P,B),(l'_P,B)}(-k_X)& -h^*_{(l_P,B),(l'_P,A)}(-k_X) & -h^*_{(l_P,B),(l'_P,B)}(-k_X) 
\end{array}\right)\nonumber
\eea
\end{widetext}
with the modified pairing coefficient
\bea
&&\bar{g}_{(l_P,s),(l'_P,s')}(k_X)\nl
&&\equiv\frac{1}{2}\Big[g_{(l_P,s),(l'_P,s')}(k_X)+g_{(l'_P,s'),(l_P,s)}(-k_X)\Big]. 
\eea
By extending the definition of operator $\tau_x$ in Eq.~(\ref{taux}) to all the matrix elements of $ D(k_X)$ labelled by ${(l_P,l'_P)}$ and using the property $\bar{g}_{(l_P,s),(l'_P,s')}(k_X)=\bar{g}_{(l_P,s),(l'_P,s')}(-k_X)$, the dynamical matrix $ D_{(l_P,l'_P)}(k_X)$ has \textit{particle-hole symmetry} expressed by
\bea
\Xi  D_{(l_P,l'_P)}(k_X) \Xi^{-1}=- D_{(l_P,l'_P)}(-k_X),
\eea
where the particle-hole operator is defined via $\Xi=\tau_xK$ satisfies $\Xi^2=+1$. 
Again, we should diagonalise the dynamical matrix $D(k_X)$, instead of the Hamiltonian, to solve EOM and obtain the eigenmodes.

\section{Experimental Conditions}\label{App:EP}


\subsection{Quasi-1D trapping potential}

In order to create a quasi-1D harmonic potential for the cold atoms, one can start from a small Bose-Einstein condensate (BEC) in a magnetic trap \cite{Paredes2004nature}. Then, the BEC is loaded into a 2D optical potential, along $y$ and $z$ directions as shown by Fig.~1(b) in the main text, by superimposing two orthogonal standing waves on top of the BEC. Each standing wave is formed by two counter-propagating Gaussian laser beams. Supposing the laser light has wavelength (wave vector) $\lambda_L$ ($k_L=2\pi/\lambda_L$), the lattice potential has the form of $V(x,y,z)=V_0(\sin^2ky+\sin^2kz)$ with the potential depth $V_0$ laser intensity. As a result, an array of 1D quantum gases confined to narrow potential tubes is created. For a sufficiently strong potential depth (laser intensity), the tunnel coupling and particle exchange between different tubes are exponentially suppressed \cite{Bloch2008RMP}. The Gaussian profile of the laser beams also leads to axial confinement of the quasi-1D gases. The resulting transverse (in the $y$-$z$ plane) trapping frequency $\omega_{tr}$ and axial trapping frequency $\omega_{ax}$ are given by \cite{Moritz2003prl}
\bea
\omega_{tr}=\frac{2E_r}{\hbar}\sqrt{\frac{V_0}{E_r}},\ \ \ \omega_{ax}=\frac{\lambda_L}{\pi w_0}\omega_{tr},
\eea
where $E_r=\hbar^2k^2_L/2m$ is the recoil energy of an atom with mass $m$, and $w_0$ is the Gaussian beam waist which sets the length of 1D harmonic trap.

\subsection{Stroboscopic lattice potential}

In our stroboscopic driving scheme, we need to control the stroboscopic lattice constant, which is usually much longer than the wavelength of laser lights.
For this purpose, one can superimpose two equally polarized laser beams of wavelength $\lambda_D$
intersecting at an angle $\theta$ as shown by Fig.~1(b) in the main text. The result is a standing wave optical dipole potential with a spatial period of \cite{Hadzibabic2004prl}
\bea
d=\frac{\lambda_D}{2\sin(\theta/2)}.
\eea
As discussed in Eq.~(\ref{Hs_SM}), we need three stroboscopic lattices with the ratio of lattice constants $$d_1:d_2:d_3=\frac{1}{2}:\frac{\sqrt{3}}{4}:\frac{\sqrt{3}}{2}$$ to create the honeycomb lattice. This can be achieved by either adjusting the angle $\theta$ of the same laser light or choosing three laser lights with different wavelengths $\lambda_D$.

\subsection{Interaction}\label{App:EP-Int}

At the low temperature in ultracold atom experiments, the collision of cold atoms is dominated by $s$-wave scattering process. The two-body interactions of ultracold gases in 3D can be described by a  pseudopotential in the form of contact function \cite{Bloch2008RMP}
\bea
V_{3D}(\tb r)=\frac{4\pi\hbar^2a}{2m}\delta(\tb r),
\eea
where $a$ is the $s$-wave scattering length. The scattering length $a$ can be further tuned by the Feshbach resonance with a magnetic field $B$, i.e.,
\bea\label{eq-sm-aB}
a(B)=a_{bg}\big[1-\frac{\Delta B}{B-B_0}\big].
\eea
Here, $a_{bg}$ is the off-resonant background scattering length, while $\Delta B$ and $B_0$ describe the width and position of the resonance. 

In the quasi-1D trap, the strength of contact interaction can be modified by the transverse mode. The effective pesodupotential is described by an interaction of the form \cite{Bloch2008RMP}
\bea
V_{1D}(x)=\frac{2\hbar\omega_{tr}a}{1-Aa/l_{tr}}\delta(x)\approx 2\hbar\omega_{tr}a\delta(x),
\eea
where the constant $A=1.036$ and $l_{tr}=\sqrt{\hbar/m\omega_{tr}}$ is the characteristic length of transverse motion. The approximation comes from the fact that the scattering length $a$ is usually much shorter than the trapping length $l_{tr}$.

\subsection{System size}\label{App:EP-SS}

During the collision, the kinetic energy of two atoms should not excite the transverse mode. This sets a restriction for the size (radius $R$) of phase space crystal, 
\bea\label{}
&&2\times\frac{1}{2}m\omega^2_{ax}\big(\frac{d}{2\pi}R\big)^2<\hbar\omega_{tr}
\eea
which results in the condition
\bea\label{eq-sm-size}
R<\frac{2\pi}{d}\sqrt{\frac{\hbar\omega_{tr}}{m\omega^2_{ax}}}=\frac{2\pi}{d}\sqrt{\frac{\hbar\pi w_0}{m\lambda_L\omega_{ax}}}.
\eea


%

\end{document}